%% file: main.tex
\def\llncs{0}
\def\anonymous{0}
\def\comments{0}

\ifnum\llncs=1
\documentclass{llncs}
\usepackage{amsmath,amssymb}
\else
\documentclass[11pt]{article}
\usepackage{amsmath,amssymb,amsthm,amsfonts}
\fi

\usepackage{mathrsfs}
\usepackage{enumitem}
\ifnum\llncs=0
\usepackage[a4paper,margin=1in]{geometry}
\fi
\usepackage{braket}
\usepackage{xcolor}

\usepackage{header}

\def\ket#1{\mathinner{|{#1}\rangle}}

\def\ketbra#1#2{\mathinner{|{#1}\rangle\!\langle{#2}|}}

\usepackage[dvipsnames]{xcolor}
\ifnum\comments=0
\newcommand{\ynote}[1]{%
  \textcolor{purple}{\textbf{[Yiming:} #1\textbf{]}}%
}

\else
\newcommand{\ynote}[1]{}
\fi
\newcommand{\qcma}{\mathsf{QCMA}}

\newcommand{\Var}{\operatorname*{Var}}

\newcommand{\dt}{\operatorname{DT}}
\newcommand{\qma}{\mathsf{QMA}}
\newcommand{\pqcma}{\mathsf{PromiseQCMA}}
\newcommand{\pqma}{\mathsf{PromiseQMA}}
\newcommand{\qcmah}{\mathsf{QCMAH}}
\newcommand{\pqcmah}{\mathsf{PromiseQCMAH}}
\newcommand{\pqmah}{\mathsf{PromiseQMAH}}
\newcommand{\bqp}{\mathsf{BQP}}
\newcommand{\ph}{\mathsf{PH}}

\newcommand{\adv}{\mathsf{adv}}
\newcommand{\byz}{\mathsf{BiasedYZ}}

\newcommand{\mult}{\mathsf{Mult}}

\usepackage{authblk}
\usepackage{orcidlink}

\ifnum\llncs=0
\usepackage{titling}
\predate{}
\postdate{}
\fi

\title{Separating $\qma$ from the $\qcma$ Hierarchy via a Switching Lemma for Polynomials}
\ifnum\anonymous=0
\author[1]{Yiming Wang}

\affil[1]{University of Washington}
\affil[ ]{\texttt{ymwang25@cs.washington.edu}}

\ifnum\llncs=0
\date{}
\fi

\else
\author{}
\institute{}
\fi
\begin{document}
\pagestyle{plain}
\maketitle
\thispagestyle{plain}

\begin{abstract}
    We construct a classical oracle relative to which $\qma$ is not contained in the $\qcma$ hierarchy. Our main technical contribution is a switching lemma for bounded polynomials, which provides a new conceptual viewpoint on earlier sampler-based arguments separating $\qma$ and $\qcma$. Crucially, the parameters of this switching lemma are independent of the number of input variables. The separation builds on the code intersection subset size problem of Bostanci, Huang, and Vaikuntanathan (FOCS 2026). We also construct a classical oracle relative to which $\qcma$ collapses to $\mathsf{P}$ but $\qma$ does not, and another relative to which $\qma$ collapses to $\mathsf{P}$ but $\mathsf{PP}$ does not.
\end{abstract}

\medskip
\medskip
\medskip
\medskip
\begin{center}
\setlength{\fboxsep}{8pt}
\fbox{%
  \parbox{0.90\textwidth}{%
    \small
    \textbf{AI usage statement.} The author used GPT-6 Astra throughout the writing of this paper, including proof techniques, exploration of related works, and polishing the language. However, the human author carried out a substantial part of this work independently. The author verified the correctness and originality of all content, including references. The author takes full responsibility for this paper. For more details, see Section~\ref{sec:ai-disclaimer}.
  }%
}
\end{center}

\ifnum\llncs=0
\newpage
\tableofcontents
\newpage
\fi
\section{Introduction}\label{sec:intro}
\input{Intro/intro}

\section{Technical Overview}
\input{Intro/tech-overview}

\section{Discussion}\label{sec:discussions}
\input{Intro/discussions}

\section{Preliminaries}
\input{Proof/prelim}

\section{Switching Lemma and $\qcmah$}\label{sec:switching}
\input{Proof/quantum-switching}

\section{Proof of the Separation}\label{sec:proof-main}
\input{Proof/proof-main}

\section{Finely Controlled Collapse of $\qcma$ and $\qma$}\label{sec:collapse}
\input{Proof/collapse}

\ifnum\llncs=1
\bibliographystyle{splncs04}
\else
\bibliographystyle{alpha}
\fi
\bibliography{main}

\appendix

\section{Black-box Definitions Inherited from \cite{BHV26}}
\input{Appendix/blackbox}

\section{AI Disclaimer}\label{sec:ai-disclaimer}
\input{Appendix/AI-disclaimer}

\end{document}

%% file: Intro/intro.tex
One of the fundamental tasks in quantum complexity theory is to understand the distinction between quantum and classical proofs. This distinction is reflected in two natural generalizations of $\mathsf{NP}$, namely $\qma$ (problems that can be efficiently verified by a quantum computer with a \emph{quantum} witness) and $\qcma$ (problems that can be efficiently verified by a quantum computer with a \emph{classical} witness). Comparing $\qma$ and $\qcma$ amounts to asking:
\begin{center}
    \textit{Are quantum proofs more powerful than classical ones?}
\end{center}
This question dates back to the work by Aharonov and Naveh~\cite{AN02}, in which $\qcma$ was first formally defined. Aaronson and Kuperberg~\cite{AK07} gave the first oracle evidence that quantum proofs may be more powerful than classical ones. They constructed a \emph{quantum} oracle relative to which some problems lie in $\qma$ but not in $\qcma$. Since then, several works~\cite{FK18,LLPY24,NN24,BD24,Zha25,LMY25} have made progress on this problem by establishing conditional or nonstandard oracle separations.

Recent work has provided stronger oracle evidence that $\qma$ strictly contains $\qcma$. Bostanci, Haferkamp, Nirkhe, and Zhandry~\cite{BHNZ26} proved the first unconditional \emph{classical} oracle separation between $\qma$ and $\qcma$. Their proof builds on spectral forrelation, a variant of the forrelation problem studied by Aaronson~\cite{Aar10} as a candidate for separating $\mathsf{BQP}$ from $\mathsf{PH}$ relative to an oracle. Zhandry~\cite{Zha25} first proposed using spectral forrelation to separate quantum and classical proofs. More recently, Bostanci, Huang, and Vaikuntanathan~\cite{BHV26} gave a simpler proof using a variant of the Yamakawa--Zhandry code intersection problem~\cite{YZ22,DQI}, building on earlier work~\cite{Liu23,LLPY24,BD24}.

These two separations~\cite{BHNZ26,BHV26}, however, share a limitation. Both use the set size estimation framework of Fefferman and Kimmel~\cite{FK18}, so in each case the separating problem lies in $\mathsf{AM}$ relative to the corresponding oracle. Each problem can therefore be decided with a single round of \emph{classical} interaction. Moreover, since the containment $\mathsf{AM}\subseteq\mathsf{BPP}^{\mathsf{NP}}$ relativizes, each separating problem has a randomized polynomial-time Turing reduction to an $\mathsf{NP}$ language, membership in which admits \emph{classical} proofs verifiable by a \emph{classical} computer. These separations therefore leave open the possibility that $\qma$ strictly contains $\qcma$, yet every language in $\qma$ can be efficiently reduced to one that admits classical proofs. This motivates the following stronger version of our original question:
\begin{center}
    \textit{Are quantum proofs \emph{truly} more powerful than classical ones?}
\end{center}

Recent work by Cakan~\cite{Cak26-QCIP} addresses this question by constructing a classical oracle separation of $\qma$ from $\mathsf{QCIP}$\footnote{The class $\mathsf{QCIP}$ is a quantum--classical generalization of $\mathsf{IP}$ in which an efficient \emph{quantum} verifier exchanges \emph{classical} messages with the prover.}. Given the inclusion $\qma\subseteq\mathsf{PSPACE}=\mathsf{IP}$~\cite{Shamir90,MW04}, one may wonder how such a separation is possible. The reason is that $\mathsf{IP}=\mathsf{PSPACE}$ does not relativize. Indeed, Fortnow and Sipser~\cite{FS88} constructed an oracle relative to which $\mathsf{coNP}\nsubseteq\mathsf{IP}$, which means $\mathsf{IP}$ lacks the power of expressing $\forall$ quantifiers in the relativized setting. Cakan's construction can be adapted so that the separating language lies in $\mathsf{coNP}$ and hence admits \emph{classical} proofs of \emph{nonmembership}.

So what does it mean for a problem to \emph{truly} require a quantum proof? We think it must satisfy at least the following two conditions. 
\begin{itemize}
    \item The problem can be efficiently verified with a \emph{quantum} witness.
    \item The problem cannot be easily \emph{reduced} to a problem for which classical proofs suffice.
\end{itemize}

How can we formalize these conditions? As an example, we may consider the class of problems that can be Turing reduced to a \(\qcma\) problem by an efficient quantum machine. Denoting this class by $\bqp^{\qcma}$, we may ask whether such reductions can eliminate the need for quantum witnesses. In other words, we ask
\[
\qma \stackrel{?}{\subseteq} \bqp^{\qcma}\,.
\]
We can be even more ambitious in asking what it means to \emph{truly} separate quantum and classical proofs. Recall that the polynomial hierarchy $\ph$ can be defined in two equivalent ways, either through a constant number of alternating quantifiers or through a tower of $\mathsf{NP}$ oracle machines. From the latter definition, we may view $\ph$ as arising from iterated $\mathsf{NP}$ oracle reductions: at each successive level, an efficient classical verifier receives a \emph{classical} witness and checks it by performing a Turing reduction to a language at the preceding level.

This motivates our focus on $\qcmah$, a quantum--classical generalization of $\ph$ first considered in~\cite{AB26}. Informally, $\qcmah$ is defined by
\[
\qcmah := \qcma \cup \qcma^{\qcma} \cup \qcma^{\qcma^{\qcma}} \cup \cdots\,.
\]
We view $\qcmah$ as capturing the full power of classical proofs by iterated $\qcma$ oracle reductions. This leads us to ask:
\begin{center}
\textit{Is $\qma$ contained in $\qcmah$?}
\end{center}
If we could construct a classical oracle relative to which $\qma$ is not contained in $\qcmah$, we would address several limitations of the earlier separations. Such a result would recover the separation of $\qma$ from $\qcma$ established in~\cite{BHNZ26,BHV26}. Moreover, the separating language would lie outside both $\ph$ and $\bqp^{\qcma}$, and hence also outside $\mathsf{BPP}^{\mathsf{NP}}$ and $\mathsf{coNP}$.

Another reason for asking this question comes from quantum cryptography, and specifically, quantum-native cryptographic primitives that may survive even if classical cryptography breaks down. Building on earlier work~\cite{KQST23}, Kretschmer, Qian, and Tal~\cite{KQT25} constructed a classical oracle relative to which $\mathsf{P}=\mathsf{NP}$ (a setting that rules out most classical cryptographic primitives) but quantum-computable one-way functions exist alongside a variety of other quantum-native cryptographic primitives. The relationship among these quantum-native primitives, however, is less well understood. Kretschmer~\cite{Kre21} constructed a \emph{quantum} oracle relative to which $\mathsf{PromiseBQP}=\pqma$, which rules out quantum-computable one-way functions, yet quantum pseudorandomness still exists relative to this oracle. Subsequent work has developed many \emph{quantum} oracle techniques for separating quantum cryptographic primitives~\cite{CCS25,BMM+25,CountCrypt,BHMV25,CCL+26,CWX26}. So far, however, these techniques have yielded only \emph{quantum} oracle separations. This leads us to ask:
\begin{center}
    \textit{Is there a \emph{classical} oracle separation between quantum-native cryptographic primitives?}
\end{center}

A natural approach is to construct a classical oracle relative to which some form of quantum cryptography exists even though $\mathsf{PromiseBQP}=\pqcma$, which would rule out quantum-computable one-way functions and several other primitives, e.g.\ efficiently verifiable one-way puzzles~\cite{CGG-EV-OWPuzz}. As a step toward this goal, we seek a classical oracle that collapses $\qcma$ while preserving as much of the power of other complexity classes as possible. This leads us to the following question:
\begin{center}
\textit{Can we build a classical oracle relative to which $\mathsf{QCMA}$ is weak but $\mathsf{QMA}$ is not?}
\end{center}
Collapsing $\qcma$ to $\mathsf{BQP}$ also collapses $\qcmah$. Keeping $\qma$ strictly more powerful therefore requires separating it from $\qcmah$, bringing us back to our earlier question.

\subsection{Our results}
We answer the complexity-theoretic questions above with the following theorems.
\begin{theorem}[Informal version of Theorem~\ref{thm:separation}]\label{thm:informal-separation}
    There exists a classical oracle $\mathcal O$ such that $$\qma^{\mathcal O}\nsubseteq \qcmah^{\mathcal O}\,.$$
\end{theorem}

As a corollary of Theorem~\ref{thm:informal-separation}, the same oracle also separates $\qma$ from other quantum--classical generalizations of \(\ph\) contained in \(\qcmah\), including the class $\mathsf{QCPH}$ considered in~\cite{GSS+18}.

We further strengthen Theorem~\ref{thm:informal-separation} by constructing an oracle relative to which \(\qcma\) collapses to \(\mathsf{P}\) while \(\qma\) remains strictly larger.

\begin{theorem}[Informal version of Theorem~\ref{thm:collapse}]\label{thm:informal-collapse}
    There exists a classical oracle $\mathcal O$ such that
    $$\mathsf{PromiseP}^{\mathcal O}=\mathsf{PromiseQCMA}^{\mathcal O}\subsetneq \mathsf{PromiseQMA}^{\mathcal O}\,.$$
\end{theorem}

The same technique also answers an open question posed in~\cite{AIK22}.

\begin{theorem}[Informal version of Theorem~\ref{thm:collapse-qma}]\label{thm:informal-collapse-qma}
    There exists a classical oracle $\mathcal O$ such that
    $$\mathsf{PromiseP}^{\mathcal O}=\mathsf{PromiseQMA}^{\mathcal O}\subsetneq \mathsf{PromisePP}^{\mathcal O}\,.$$
\end{theorem}

We note that the separating problems in Theorems~\ref{thm:informal-collapse}~and~\ref{thm:informal-collapse-qma} can both be chosen to be decision problems (as opposed to promise problems).

\subsection{Technical contributions}
While Theorem~\ref{thm:informal-separation} is the main result of this paper, we believe its proof is of independent interest. A central ingredient is the following lemma, which we call the \emph{polynomial switching lemma}.

\begin{lemma}[Informal version of Theorem~\ref{thm:quantum_switching}]\label{lem:informal_switching}
    Let $f$ be a bounded polynomial of degree $T$, and let $\rho$ be a random restriction with $\Pr[*]=p$. Then, on the restricted subcube defined by $\rho$, $f_\rho$ can be $\eta$-approximated by a classical deterministic decision tree of depth less than $t$, with probability at least $1-\left(O(pT^2/\eta)\right)^t$.
\end{lemma}

Crucially, the bound in Lemma~\ref{lem:informal_switching} does \emph{not} depend on the number of variables in $f$. For comparison, we recall the following H{\aa}stad switching lemma.\footnote{Despite the similarity, they are incomparable: the width of a DNF or CNF seems unrelated to the degree of the corresponding polynomial.}

\begin{lemma}[\cite{Hastad86}]
    Let $f$ be a DNF (or CNF) of width $w$, and let $\rho$ be a random restriction with $\Pr[*]=p$. Then, on the restricted subcube defined by $\rho$, $f_\rho$ can be computed by a classical deterministic decision tree of depth less than $t$, with probability at least $1-\big(O(pw)\big)^t$.
\end{lemma}

We believe that Lemma~\ref{lem:informal_switching} and its proof abstract and generalize the sampler-based argument of Fefferman and Kimmel~\cite{FK18}, which underlies the recent classical oracle separations of $\qma$ from $\qcma$~\cite{BHNZ26,BHV26}. Identifying the role of restrictions in this argument provides a new conceptual viewpoint and leads to an alternative, arguably clearer proof of the separation. We discuss the lemma's connections to several related lines of research below.

\paragraph{Switching lemmas.}
The switching lemma originated in the study of the computational limits of $\mathsf{AC}^0$ circuits. H{\aa}stad~\cite{Hastad86} used his switching lemma to prove sharp lower bounds on the size of $\mathsf{AC}^0$ circuits computing parity. Since then, the lemma and its variants have found many applications. In particular, they have been used to show that $\mathsf{AC}^0$ circuits can predict parity with only a negligible advantage over random guessing~\cite{Hastad14a}; to prove that the polynomial hierarchy is infinite with probability one relative to a random oracle~\cite{RST15}; to prove that the Fourier mass of $\mathsf{AC}^0$ functions is concentrated at low degrees~\cite{LMN89,Tal17}; and to construct pseudorandom generators that fool $\mathsf{AC}^0$ circuits~\cite{TX13,ST19,Lyu22}.

Several quantum analogues of the switching lemma were already known. Aaronson, Ingram, and Kretschmer~\cite{AIK22} proved a version for $\qma$ query algorithms. More recently, Agarwal and Ben-David~\cite{AB26} proved a more closely related switching lemma for bounded polynomials, with applications to $\bqp$ query algorithms and the $\qcma$ hierarchy. In both results, however, the parameters depend on the number of input variables, while in ours they do not.

Our polynomial switching lemma (Lemma~\ref{lem:informal_switching}) does not directly extend to $\qma$ query algorithms: if it did, our lower-bound argument would also apply to the $\qma$ algorithm for our separating problem. Our lemma does, however, subsume the switching lemma of~\cite{AB26}. We expect that an adaptation to random projections would also subsume their random projection lemma.

We hope that our polynomial switching lemma (Lemma~\ref{lem:informal_switching}) will shed light on quantum query algorithms. Much as the H{\aa}stad switching lemma did in the classical setting, our lemma may inspire new techniques in the study of constant-depth quantum circuits and even beyond.

\paragraph{Aaronson--Ambainis conjecture.}
Our polynomial switching lemma also connects to the Aaronson--Ambainis conjecture\footnote{Turturean~\cite{AAproof} recently proved the conjecture. Here, we retain its historical name.}, which is widely known in the following form.

\begin{conjecture}[\cite{AA11}]\label{conj:aaronson-ambainis}
    Let $f$ be a bounded polynomial of degree $T$. Then $f$ has a highly influential variable whose influence is at least $\poly(\Var(f),1/T)$.
\end{conjecture}

How does this conjecture relate to our switching lemma? Aaronson and Ambainis~\cite{AA11} showed that Conjecture~\ref{conj:aaronson-ambainis} implies the following result.

\begin{conjecture}\label{conj:AA-folklore}
    Let $\epsilon,\delta>0$. Any bounded polynomial of degree $T$ can be $\epsilon$-approximated by a classical deterministic decision tree of depth $\poly(T,1/\epsilon,1/\delta)$ on a $1-\delta$ fraction of the inputs.
\end{conjecture}

Our polynomial switching lemma (Lemma~\ref{lem:informal_switching}) can be viewed as a version of Conjecture~\ref{conj:AA-folklore} under random restrictions. As in the conjecture, our decision tree depth bound is independent of the number of input variables. However, the decision tree in our lemma approximates the polynomial on \emph{all} inputs in the restricted subcube, whereas Conjecture~\ref{conj:AA-folklore} considers approximation on \emph{most} of the inputs. This distinction turns out to be valuable in our proof separating \(\qma\) from \(\qcmah\).

Earlier efforts to prove the Aaronson--Ambainis conjecture (Conjecture~\ref{conj:aaronson-ambainis}) yielded a series of partial results~\cite{OZ16,ABP18,BSdW,Gut23,LZ23,Bha25}. More recently, Turturean~\cite{AAproof} gave a Lean-verified proof of the conjecture, thereby also establishing Conjecture~\ref{conj:AA-folklore}. We hope that our polynomial switching lemma (Lemma~\ref{lem:informal_switching}) and its proof might shed light on this proof and potentially suggest a way to strengthen Conjecture~\ref{conj:AA-folklore}.


\paragraph{Related work.}
Lovett and Zhang~\cite{LZ23} proposed an approach to the Aaronson--Ambainis conjecture using fractional block sensitivity, which also played a role in the eventual proof~\cite{AAproof}. Fractional block sensitivity appears both in the proof of a weaker switching lemma~\cite{AB26} and in our own proof. We wonder whether these proofs share a common idea and whether our switching lemma might help simplify the proof of the Aaronson--Ambainis conjecture.

Bhattacharya~\cite{Bha25} proved a different version of the Aaronson--Ambainis conjecture under random restrictions. Their result concerns approximation on \emph{most} inputs of the restricted subcube, while ours gives a bound for approximation on \emph{all} such inputs. The two results are therefore incomparable.

Liu, Mutreja, and Yuen~\cite{LMY25} also explored a connection between the $\qma$ versus $\qcma$ separation and the Aaronson--Ambainis conjecture. Assuming a conjecture closely related to the Aaronson--Ambainis conjecture, they constructed a classical oracle separating $\qma$ from $\qcma$. Our connection to the Aaronson--Ambainis conjecture arises through our polynomial switching lemma, which is different from their proposed route.

\paragraph{Concurrent work.} Cakan~\cite{Cak26-QCMAH} also constructed a classical oracle separating $\qma$ from $\qcmah$, and a second classical oracle separating pseudorandom quantum states from quantum-computable one-way functions. Our work is perhaps conceptually simpler and takes a different direction by proving the polynomial switching lemma (Lemma~\ref{lem:informal_switching}), which is of independent interest. Cakan's parameter choices also differ from ours.

In another paper, Cakan~\cite{Cak26-QCIP} separated $\mathsf{QMA}$ from $\mathsf{QCIP}$. The paper appeared online shortly after we obtained an initial version of our $\qma$--$\qcmah$ separation (see Section~\ref{sec:ai-disclaimer}). As discussed in Section~\ref{sec:intro}, the two separations are incomparable.

Finally, we note that the recent proof of the Aaronson--Ambainis conjecture~\cite{AAproof,BDST26} does not subsume our polynomial switching lemma.

%% file: Intro/tech-overview.tex
\paragraph{Code intersection subset size problem.} We begin by reviewing the proof strategy of Bostanci, Huang, and Vaikuntanathan~\cite{BHV26}. As discussed in Section~\ref{sec:intro}, this strategy has its roots in the set size estimation framework of Fefferman and Kimmel~\cite{FK18}. The same framework serves as the starting point for our polynomial switching lemma.

At a high level, the separating problem in~\cite{BHV26} is as follows.

\begin{tcolorbox}[
    breakable,
    colback=white,
    colframe=black,
    boxrule=0.5pt,
    arc=0pt,
    left=8pt, right=8pt, top=8pt, bottom=8pt
]
        \begin{center}
            \textbf{The code intersection subset size problem}
        \end{center}
        \textbf{Parameters:}
        \begin{itemize}
            \item A sufficiently large positive integer $\lambda$, the security parameter, given in unary as $1^\lambda$.
            \item A set $A$ satisfying $|A|=2^{\poly(\lambda)}\gg 2^{2\lambda}$.
            \item A distribution $\mathcal D$ over subsets $D_H\subseteq A$ of size at least $2^\lambda$.
        \end{itemize}
        \textbf{Problem:} Given black-box access to a set $O\subseteq A$\footnotemark{} drawn from one of the following two distributions, determine which distribution generated $O$.
        \begin{itemize}
            \item ``YES'' distribution: $O$ is sampled from $\mathcal D$. In particular, $|O|\geq 2^\lambda$.
            \item ``NO'' distribution: $O$ is sampled from subsets of size $|O|\leq 2^\lambda/2$.
        \end{itemize}
        \textbf{Guarantees:}
        \begin{enumerate}
            \item There is a quantum algorithm $\mathsf{Decode}$ that takes $1^\lambda$ and a quantum state as input, runs in $\poly(\lambda)$ time, and outputs a point in $A$. It has the following properties.
            \begin{itemize}
                \item With high probability over $D_H\gets\mathcal D$, there exists a quantum witness $\ket{\adv_H}$ such that $\mathsf{Decode}(\ket{\adv_H})$ outputs a point in $D_H$ with high probability.
                \item For any witness $\rho$ and any $x\in A$, $\Pr[\mathsf{Decode}(\rho)=x]\leq 2^{-\lambda}$.
            \end{itemize}

            \item There is a constant $p\in (0,1)$, independent of $\lambda$, such that for every $S\subseteq A$ with $|S|\leq 2^\lambda$,
            \[
                \Pr_{D_H\gets\mathcal D}\left[S\subseteq D_H\right]\leq p^{\lambda\cdot |S|}\,.
            \]
        \end{enumerate}
\end{tcolorbox}
\footnotetext{Throughout, we identify $O$ with its membership function $\mathbf{1}_O$.}

This abstract description is almost all we need. Taking the guarantees as a black box, we can show that the \emph{code intersection subset size} problem lies in $\qma\setminus\qcma$.

Guarantee (a) gives us the $\qma$ verifier directly. The verifier simply runs $\mathsf{Decode}$ on the quantum witness $\ket{w}$ and accepts if the output belongs to $O$.

How do we show that the problem lies outside $\qcma$? We compare it to the \emph{set size estimation} problem, for which Fefferman and Kimmel~\cite{FK18} give an elegant proof of nonmembership in $\qcma$. A similar argument shows that the code intersection subset size problem lies outside $\qcma$, recovering the proof of~\cite{BHV26}. We next review the argument of~\cite{FK18}, whose central idea also underlies our polynomial switching lemma.

\paragraph{Set size estimation.} As discussed above, we use Guarantee (a) only to construct a $\qma$ verifier. Setting this guarantee aside, we can see the similarity between the \emph{code intersection subset size} problem and the following \emph{set size estimation} problem.

\begin{tcolorbox}[
    breakable,
    colback=white,
    colframe=black,
    boxrule=0.5pt,
    arc=0pt,
    left=8pt, right=8pt, top=8pt, bottom=8pt
]
        \begin{center}
            \textbf{The set size estimation problem}
        \end{center}
        \textbf{Parameters:}
        \begin{itemize}
            \item A sufficiently large positive integer $\lambda$, the security parameter, given in unary as $1^\lambda$.
            \item A set $A$ satisfying $|A|=2^{\poly(\lambda)}\gg 2^{2\lambda}$.
        \end{itemize}
        \textbf{Problem:} Given black-box access to a set $O\subseteq A$ drawn from one of the following two distributions, determine which distribution generated $O$.
        \begin{itemize}
            \item ``YES'' distribution: $O$ is sampled uniformly from subsets of size $|O|=2^\lambda$.
            \item ``NO'' distribution: $O$ is sampled from subsets of size $|O|\leq 2^\lambda/2$.
        \end{itemize}
\end{tcolorbox}

Our goal is to show that the \emph{set size estimation} problem lies outside $\qcma$. We first review the proof of~\cite{FK18} and explain how it applies to the \emph{code intersection subset size} problem. We then give our interpretation of the argument.

The intuition behind~\cite{FK18} is simple. For a ``YES'' instance, the proof should encode enough information to identify more than $2^\lambda/2$ points in $O$, which suggests that the $\qcma$ witness must be exponentially long. We formalize this intuition using the extractor in Algorithm~\ref{alg:extractor}.

\begin{algorithm}[htbp]
    \caption{Extractor $\mathsf{Ext}$}
    \label{alg:extractor}
    \begin{algorithmic}
        \State \textbf{Input:} The $\qcma$ verifier $V$ and the witness $w$
        \State Set $S_0:=\varnothing$
        \State Let $Q$ be the number of oracle queries made by $V^{(\cdot)}(1^\lambda,w)$
        \For{$i=1,2,\ldots,2^\lambda/2$}
            \State Sample $t$ uniformly from $[Q]$
            \State Simulate $V^{S_{i-1}}(1^\lambda,w)$ until just before its $t$-th oracle query
            \State Measure the query address register and let $x_i$ be the outcome
            \State Set $S_i:=S_{i-1}\cup\{x_i\}$
        \EndFor
        \State \textbf{Output:} $S_{2^\lambda/2}$
    \end{algorithmic}
\end{algorithm}

\paragraph{The probability of successful extraction.}
Suppose $O$ is a ``YES'' instance, so that $|O|=2^\lambda$, and let $w$ be a $\qcma$ witness for $O$.
For a fixed verifier $V$, the extractor receives only $w$ and has no access to $O$. Even so, we claim that it outputs $2^\lambda/2$ distinct points in $O$ with much higher probability than random guessing would allow.
How can we obtain a lower bound on this probability, and how does that bound lead to a contradiction?

Suppose that the witness $w$ for $O$ is known. Consider the first iteration ($i=1$). The oracles $\mathbf{1}_O$ and $\mathbf{1}_\varnothing$ differ only on $O$. Since $O$ is a ``YES'' instance with witness $w$ and $\varnothing$ is a ``NO'' instance, the acceptance probabilities of $V^O(1^\lambda,w)$ and $V^\varnothing(1^\lambda,w)$ differ by at least a constant. The algorithm $V^{\varnothing}(1^\lambda,w)$ must therefore place significant query weight on $O$. Measuring the query register thus yields $x_1\in O$ with probability at least $1/\poly(Q)$, and hence at least $1/\poly(\lambda)$.

Now consider the second iteration ($i=2$), conditioned on the success of the first iteration. The set $S_1=\{x_1\}$ is still a ``NO'' instance, and $\mathbf{1}_{S_1}$ and $\mathbf{1}_O$ differ only on $O\setminus\{x_1\}$. By the same argument, $V^{S_1}(1^\lambda,w)$ places significant query weight on $O\setminus\{x_1\}$. Thus the probability of extracting a second distinct point satisfies
\[
    \Pr\left[x_2\in O\backslash\{x_1\}\mid x_1\in O\right]\geq 1/\poly(\lambda)\,.
\]

We can continue this analysis through iteration $i=2^\lambda/2$, since $S_{i-1}$ is still a ``NO'' instance at each step. Multiplying the probabilities of extracting a new point in $O$, we obtain
\begin{equation*}
    \Pr_{S\gets \mathsf{Ext}(V,w)}[S\subseteq O\land |S|=2^\lambda/2]
    \geq \left(\frac{1}{\poly(\lambda)}\right)^{2^\lambda/2}\,.
\end{equation*}
If $w$ is unknown, we may simply guess one.
For every ``YES'' instance $O$, the probability of choosing a valid witness is at least $2^{-|w|}$. Thus,
\begin{equation}\label{eq:tech-overview-extractor}
    \Pr_{\substack{w\gets \{0,1\}^{|w|}\\S\gets \mathsf{Ext}(V,w)}}
    [S\subseteq O\land |S|=2^\lambda/2]
    \geq 2^{-|w|}\left(\frac{1}{\poly(\lambda)}\right)^{2^\lambda/2}\,.
\end{equation}
In Equation~\eqref{eq:tech-overview-extractor}, note that the distribution of $S$ is completely independent of $O$.

On the other hand, for any fixed $S$ of size $2^\lambda/2$, a uniformly random ``YES'' instance $O$ satisfies
\begin{equation}\label{eq:tech-overview-bound}
    \Pr[S\subseteq O]\leq \left(\frac{2^\lambda}{|A|}\right)^{|S|}
    \ll 2^{-\lambda\cdot |S|}\,.
\end{equation}
If $|w|=\poly(\lambda)$, comparing Equation~\eqref{eq:tech-overview-extractor} and Equation~\eqref{eq:tech-overview-bound} gives the desired contradiction.

As a final note, using Guarantee (b) in place of Equation~\eqref{eq:tech-overview-bound} provides the argument for the \emph{code intersection subset size} problem of~\cite{BHV26}.

\paragraph{What are the essential ingredients?} This method may apply well beyond the problems considered above. How might we apply it to other problems? To answer this question, we recast the argument in more general terms to identify the assumptions on which the proof relies.

\subparagraph{Observation 1.} The proof makes little direct use of the fact that the witness is classical or that it can be cloned.\footnote{As pointed out in~\cite{CCHY-cloning-separation}, showing that the same problem lies outside $\mathsf{ClonableQMA}$ presents difficulties.} The extractor $\mathsf{Ext}$ (Algorithm~\ref{alg:extractor}) uses $w$ only through the computation $V^{(\cdot)}(1^\lambda,w)$. For a fixed verifier $V$, each choice of $w$ specifies a \emph{$\bqp$ query algorithm}. Once we bound the extraction probability for each such algorithm, the rest is a \emph{union bound} over the $2^{|w|}$ possible witnesses. Thus, the natural objects to study are \emph{quantum query algorithms}.

\subparagraph{Observation 2.} For which other problems does the same extractor $\mathsf{Ext}$ work? At the $i$-th iteration, we can extract a favorable $x_i$ because \emph{the current instance $S_{i-1}$ is a ``NO'' instance, and there is a ``YES'' instance that differs from it only on $O\setminus S_{i-1}$}. In particular, we do not need to know that $O$ itself is a ``YES'' instance. More generally, we do not even need to know whether the current instance is a ``YES'' or a ``NO'' instance. It suffices that \emph{there are both a ``YES'' instance and a ``NO'' instance that agree with the current instance outside $O\setminus S_{i-1}$}. The acceptance probability on at least one of these two instances must differ by a constant from that on the current instance. Thus, on the current instance, the verifier must place significant query weight on $O\setminus S_{i-1}$.

What does this observation tell us about the problem? The set $O$ is fixed, and as long as every point extracted so far is favorable, the extractor modifies the instance only inside $O$. We may therefore consider the problem \emph{restricted to} the subcube $\{0,1\}^O\times\{0\}^{A\setminus O}$. The condition for the $i$-th iteration now has a simple interpretation: \emph{even after the values at $x_1,\ldots,x_{i-1}$ have been fixed and known, the restricted subcube still contains both a ``YES'' instance and a ``NO'' instance}.

This is precisely where \emph{decision tree depth} enters the argument. Assume that the extractor can adversarially choose the values at $x_1,\ldots,x_{i-1}$. If the decision tree depth on the restricted subcube is at least $i$, the extractor can ensure that the condition for the $i$-th iteration holds (since knowing the values at any collection of $i-1$ points does not determine the output).

Finally, the ``background'' of the restricted subcube need not be all zeros. We may assign each $x\in A$ a fixed background value, as long as this assignment is known to the extractor. Thus, the extractor can still obtain a favorable point at each of the first $t$ steps if \emph{the decision tree depth on the restricted subcube is at least $t$}.

\paragraph{Set size estimation, revisited.} We now revisit the proof that the \emph{set size estimation} problem is not in $\qcma$. (For simplicity, we omit the quantitative bounds.)

\begin{enumerate}
    \item Suppose $O$ is sampled from a distribution $\mathcal D$. Equation~\eqref{eq:tech-overview-bound} (Guarantee (b)) states that a random $O$ is unlikely to contain any given set of the required size.
    \item By Observation~2, if the problem still has high decision tree depth on the restricted subcube with $O$ unrestricted, then the extractor $\mathsf{Ext}$ outputs a subset of $O$ of the required size with good probability.
    \item Combining (a) and (b), any problem that can be solved by a $\bqp$ query algorithm must, with high probability, have low decision tree depth after a random restriction drawn according to $\mathcal D$.
    \item A $\qcma$ algorithm can be viewed as an OR over $\bqp$ query algorithms. By (c) and a union bound, any problem that can be solved by a $\qcma$ algorithm must therefore reduce to a small-width DNF with high probability after a random restriction.
    \item The structure of the \emph{set size estimation} problem implies that it cannot reduce to a small-width DNF under such a restriction.
\end{enumerate}

One may notice a striking similarity between (c) and the H{\aa}stad switching lemma. Corollary~\ref{cor:quantum_switching} gives a quantitative version of (c), which is the switching lemma for quantum query algorithms.

\paragraph{Generalizing to polynomials.} Having established the switching lemma for quantum query algorithms, we may naturally ask whether it also applies to polynomials. To see why it does, we recall how Algorithm~\ref{alg:extractor} uses the quantum query algorithm.

\subparagraph{Observation 3.} Suppose a ``YES'' instance $O_1$ and a ``NO'' instance $O_2$ differ only on $D$. When the algorithm runs on $O_2$, sampling according to its average query weights yields a point in $D$ with probability $\Omega(1/Q^2)$, where $Q$ is the number of queries.

How can we generalize this observation to polynomials? This is where we introduce \emph{fractional certificates} (Definition~\ref{def:fbs}). Informally, a fractional certificate is an assignment of weights to all variables, so that modifying a set of variables changes the value of the polynomial by at most the sum of their weights. Crucially, for a polynomial $f$ taking values in $[0,1]$, these weights can be chosen so that the total weight is $O(\deg(f)^2)$ at every input. Thus, if modifying a set of variables $D$ changes the value of $f$ by at least $\epsilon$, sampling a variable with probability proportional to its weight yields a point in $D$ with probability $\Omega(\epsilon/\deg(f)^2)$. This gives the desired extractor for polynomials and hence the polynomial switching lemma (Lemma~\ref{lem:informal_switching}).

\paragraph{Application to $\qcmah$.} Given the similarity between our polynomial switching lemma and the H{\aa}stad switching lemma, it is natural to apply our lemma to constant-depth circuits. The following lemma from~\cite{AB26} provides a starting point for analyzing $\qcmah$.

\begin{lemma}[Informal version of Proposition~\ref{prop:qcmah-circuits}]\label{lem:tech-overview-qcmah-circuit}
    Any problem in $\qcmah^{\mathcal O}_d$ can be solved by a constant-depth circuit of quasipolynomial size that takes the truth table of $\mathcal O$ as input and uses only OR gates (of unbounded fan-in) and $\bqp$-query gates.\footnote{We define these circuits later in this section.}
\end{lemma}

Using the polynomial switching lemma and a generalization of the H{\aa}stad switching lemma~\cite{Kelley21}, we obtain the following theorem.

\begin{theorem}[Informal version of Theorem~\ref{thm:qcmah-restriction}]\label{thm:tech-overview-qcmah-restriction}
    Suppose we sample a family of sets $D_0\supseteq D_1\supseteq\cdots\supseteq D_d$ such that, conditioned on $D_{i-1}$, any fixed subset of $D_{i-1}$ is unlikely to be contained in $D_i$. Fix a depth-$d$ circuit of the type described in Lemma~\ref{lem:tech-overview-qcmah-circuit}. Then, a random restriction that leaves $D_d$ unrestricted and fixes the remaining variables to independent uniform bits reduces the circuit to a low-depth decision tree with high probability.
\end{theorem}

It is thus natural to apply this theorem to the following problem.

\begin{tcolorbox}[
    breakable,
    colback=white,
    colframe=black,
    boxrule=0.5pt,
    arc=0pt,
    left=8pt, right=8pt, top=8pt, bottom=8pt
]
        \begin{center}
            \textbf{Our separating problem}
        \end{center}
        \textbf{Parameters:} 
        \begin{itemize}
            \item a sufficiently large positive integer $\lambda$, the security parameter, given in unary as $1^\lambda$,
            \item a set $A$ satisfying $|A|=2^{\poly(\lambda)}\gg 2^{2\lambda}$, and
            \item a distribution $\mathcal D$ over subsets of $A$ of size at least $2^\lambda$.
        \end{itemize}
        \textbf{Problem:} Given black-box access to a Boolean function $f:A\to\{0,1\}$ drawn from one of the following two distributions, determine which distribution generated $f$.
        \begin{itemize}
            \item ``YES'' distribution: $f=R\lor\mathbf{1}_D$, where $D$ is sampled from $\mathcal D$ and $R:A\to\{0,1\}$ is a uniformly random function sampled independently of $D$.
            \item ``NO'' distribution: $f=R$, where $R:A\to\{0,1\}$ is a uniformly random function.
        \end{itemize}
        \textbf{Guarantees:}
        \begin{enumerate}
            \item The same as Guarantee (a) of the \emph{code intersection subset size} problem.

            \item There is a constant $p\in (0,1)$, independent of $\lambda$, such that for every $d\in\mathbb N_+$, there is a distribution over families of sets
            \[
                A=D_0\supseteq D_1\supseteq\cdots\supseteq D_d
            \]
            with $D_d$ identically distributed as $\mathcal D$. Furthermore, for each $i\in[d]$, each choice of $D_{i-1}$, and each $S\subseteq D_{i-1}$ with $|S|\leq 2^\lambda$, conditioned on the choice of $D_{i-1}$,
            \[
                \Pr[S\subseteq D_i]\leq p^{\lambda\cdot |S|/d}\,.
            \]
        \end{enumerate}
\end{tcolorbox}

Informally, we obtain a ``YES'' instance by taking the pointwise OR of the membership function $\mathbf{1}_D$ from the \emph{code intersection subset size} problem and an independent uniformly random function $R$. We introduce $R$ because the H{\aa}stad switching lemma that we use (Theorem~\ref{thm:kelley-switching}) requires the restricted variables to be fixed to \emph{independent, uniformly random} bits.

The construction in~\cite{BHV26} can be easily shown to satisfy Guarantee (b) above. Applying Theorem~\ref{thm:tech-overview-qcmah-restriction} then shows that our separating problem lies outside $\qcmah$.

How do we show that our separating problem lies in $\qma$? Completeness follows immediately from Guarantee (a). For soundness, it suffices to show that, with high probability over a random ``NO'' instance, the maximum acceptance probability of the verifier over all witnesses is bounded away from $1$. To see this, we need to take a closer look inside the black box. For the \cite{BHV26} verifier, the maximum acceptance probability is the spectral norm of the average over exponentially many independent random matrices, whose variance is bounded. This lets us apply a matrix concentration inequality, getting the desired soundness bound.

\paragraph{Collapsing $\qcma$ to $\mathsf{P}$.}
A classical result on oracle separations between $\mathsf{PP}$ and $\ph$ states that there exists an oracle relative to which $\mathsf{P}=\ph$ while $\mathsf{PP}$ remains strictly larger. However, it is not immediate how to construct an oracle that collapses $\qcma$ to $\mathsf{P}$ while keeping $\qma$ strictly larger. The difficulty, as noted in~\cite{AIK22}, concerns how the oracle answers the out-of-promise queries.

Informally, to collapse $\qcma$ to $\mathsf{P}$ while keeping $\qma$ strictly larger, it suffices to construct an oracle with two parts. The first provides the problem separating $\qma$ from $\qcmah$. The second collapses $\qcma$ to $\mathsf{P}$ by recording the answers of all $\qcma$ verifiers relative to the oracle. By recursively expanding queries to the second part according to the semantics, we obtain a low-depth circuit of $\qcma$-query gates.

The difficulty is that a $\qcma$ algorithm specifies no answer on out-of-promise inputs, yet the classical oracle must return either $0$ or $1$. The choice of these answers may unintentionally reveal information to a $\mathsf{P}$ algorithm. For example, if the oracle returns $1$ exactly when the maximum acceptance probability exceeds $1/2$, a $\mathsf{P}$ algorithm can decide any language in $\mathsf{PP}$, since the probability estimation of a quantum circuit is $\mathsf{PP}$-complete.

A related subtlety arises in Lemma~\ref{lem:tech-overview-qcmah-circuit}. How do we define the value of a circuit whose intermediate gates may compute partial functions? More fundamentally, how do we define, say, $\qcma^\qcma$ when the $\qcma$-complete problem is a promise problem?

We answer both questions with the following convention.

\begin{definition}[Informal version of Definition~\ref{def:partial-func-extension}]\label{def:tech-overview-convention}
    Fix an input to a circuit whose gates may compute partial functions. Define the evaluation of this circuit as follows.
    \begin{itemize}
        \item The circuit outputs $0$ if every total completion of the circuit outputs $0$ on this input.
        \item The circuit outputs $1$ if every total completion of the circuit outputs $1$ on this input.
        \item The circuit outputs $\perp$ otherwise.
    \end{itemize}
\end{definition}

How does Definition~\ref{def:tech-overview-convention} differ from the convention used in~\cite{AIK22}? The difference lies in how the output of an intermediate gate is reused. Suppose a gate $g$ outputs $\perp$ on every input and the circuit output is $g\oplus g$. Under our convention, both occurrences of $g$ have the same value in each total completion, so the circuit always outputs $0$. Under the convention in~\cite{AIK22}, the output gate receives $\perp\oplus\perp$, and the two occurrences of $\perp$ can take different values, so the circuit outputs $\perp$.

Our convention (Definition~\ref{def:tech-overview-convention}) is useful for the following two reasons.

\begin{itemize}
    \item It captures the definition of $\qcmah$ in~\cite{AIK22,AB26}. Informally, for a promise language $\Pi$, a promise problem is in $\qcma^{\Pi}$ if there is a single oracle-$\qcma$ \emph{algorithm} that decides the problem correctly on every promised input for every total completion of $\Pi$. Crucially, on a ``YES'' input, the accepting witness may depend on the completion of $\Pi$.
    Under the correspondence in Lemma~\ref{lem:tech-overview-qcmah-circuit}, Definition~\ref{def:partial-func-extension} captures precisely this definition for $\qcmah$ with oracle-dependent witnesses.

    \item To show that the separating problem lies outside $\qcmah$, it suffices to find, for each candidate circuit, an input and \emph{some} consistent assignment of values to the oracle-$\qcma$ verifiers for which the circuit gives the wrong answer. This is precisely the consistency required by Definition~\ref{def:tech-overview-convention}. On that input, each intermediate gate has a single value, which is reused wherever its output is needed.
\end{itemize}

Thus, Definition~\ref{def:tech-overview-convention} helps resolve the difficulty in constructing an oracle that collapses $\qcma$ to $\mathsf{P}$. The answers to out-of-promise queries need not be fixed \emph{a priori}. Instead, we assign them during the diagonalization. At each stage, it suffices to find \emph{one} consistent assignment to the intermediate gates under which the current machine gives the wrong answer on the chosen input. We then use that assignment to determine the corresponding oracle values.

Under Definition~\ref{def:tech-overview-convention}, the proof of Theorem~\ref{thm:tech-overview-qcmah-restriction} requires additional care. We refer the reader to the proof of Theorem~\ref{thm:qcmah-restriction} for details.

%% file: Intro/discussions.tex
\paragraph{Open problems.} Our results naturally lead to the following directions for further research.
\begin{enumerate}
    \item Can we prove the same separation for the $\mathsf{ClonableQMA}$ hierarchy? As we shall see, our proof relies on an explicit description of all possible witnesses. For a $\mathsf{ClonableQMA}$ verifier, the cloning algorithm, and hence the set of clonable witnesses, can depend on the oracle. These witnesses may also be hard to prepare. Recent work~\cite{CCHY-cloning-separation} constructed a classical oracle separating $\mathsf{ClonableQMA}$ from $\mathsf{QCMA}$ and identified a difficulty in extending the proofs of~\cite{BHNZ26,BHV26} to separate $\mathsf{QMA}$ from $\mathsf{ClonableQMA}$. It would thus be interesting to understand whether our proof technique applies to $\mathsf{ClonableQMA}$.

    \item Can we prove the separation relative to a \emph{random} oracle, or in other words, a separation without structure? This would show that quantum proofs retain their advantage even in the query model. The Aaronson--Ambainis conjecture limits the advantage of quantum over classical query \emph{algorithms} on uniformly random inputs, but it leaves open the possibility of a random oracle separation between quantum and classical \emph{proofs}.

    Our separating oracle is arguably less structured than those in previous works~\cite{BHNZ26,BHV26,Cak26-QCIP}. However, we still need to plant a valid code intersection block in each ``YES'' instance, and such a block occurs with only doubly exponentially small probability in a random oracle. Can we find a variant of this problem for which both ``YES'' and ``NO'' instances occur with good probability under a random oracle?

    Our work also gives a straightforward way to separate many $\mathsf{P}^\mathsf{\#P}$ languages from $\mathsf{QCMAH}$. Can we provide some form of ``assistance'' for natural $\mathsf{P}^\mathsf{\#P}$ problems such as parity or majority relative to a random oracle, so that they lie in $\mathsf{QMA}$ (or at least in $\mathsf{QMAH}$) while remaining outside $\mathsf{QCMAH}$?

    \item As observed in~\cite{BHV26}, it would be interesting to determine whether a similar decoding problem is $\mathsf{QMA}$-complete. Recent results~\cite{MN26,GR26-QMA-QMA1} show that the spectral forrelation problem of~\cite{BHNZ26} is $\qma$-complete if the two oracles are instead specified by classical CSPs. Can we prove that our code intersection problem is $\mathsf{QMA}$-complete when the oracle has a succinct description?

    \item Our switching lemma seems to have a deeper connection to the Aaronson--Ambainis conjecture, particularly in light of the framework introduced by Lovett and Zhang~\cite{LZ23} and the subsequent proof of the conjecture~\cite{AAproof}. Can we find a simpler proof of the conjecture using our polynomial switching lemma and connect the proof to the sampler of Fefferman and Kimmel~\cite{FK18}?
\end{enumerate}

\paragraph{Acknowledgments.}
The author thanks Andrea Coladangelo and Chinmay Nirkhe for helpful discussions.

%% file: Proof/prelim.tex
\subsection{Complexity theory}

\begin{definition}
    A (total) language is defined as a subset of $\{0,1\}^*$.

    A decision problem is a function $\mathcal L:\{0,1\}^*\to \{0,1\}$. We identify each decision problem $\mathcal L$ with the total language $\mathcal L^{-1}(1)$. Thus, we use the two notations interchangeably.
\end{definition}

\begin{definition}
    A promise language
    is defined as a pair of sets of ``YES" instances $\mathcal L_{\text{YES}}\subseteq\{0,1\}^*$ and ``NO" instances $\mathcal L_{\text{NO}}\subseteq \{0,1\}^*$, such that $\mathcal L_{\text{YES}}\cap \mathcal L_{\text{NO}}=\varnothing$.

    A promise problem is a function $\mathcal L:\{0,1\}^*\to \{0,1,\perp\}$. We identify each promise problem $\mathcal L$ with the promise language $(\mathcal L_{\text{YES}}=\mathcal L^{-1}(1),\mathcal L_{\text{NO}}=\mathcal L^{-1}(0))$. Thus, we use the two notations interchangeably.

    A decision problem $\mathcal K$ is a \emph{total completion} of the promise problem $\mathcal L$, if for any $x\in \{0,1\}^*$,
    $$\mathcal L(x)=1\implies \mathcal K(x)=1\,,\quad \mathcal L(x)=0\implies \mathcal K(x)=0\,.$$
\end{definition}


\begin{definition}
    We define a \emph{(classical) oracle} to be a decision problem $\mathcal O:\{0,1\}^*\to \{0,1\}$.

    For every parameter $\lambda\in \mathbb N_+$ and every function $f:\{0,1\}^\lambda\to \{0,1\}$, define a quantum circuit that queries $f$ as follows. Let $O_f$ be the unitary that acts as $\ket{x,b}\mapsto \ket{x,b\oplus f(x)}$. The quantum circuit is of the form
    $$U_{T+1}O_fU_TO_f\cdots O_fU_2O_fU_1$$
    where $U_1,U_2,\cdots,U_{T+1}$ are circuits that consist of elementary gates from a certain universal gate set. The integer $T$ is the \emph{number of queries} of this circuit.

    Let $\mathcal O$ be a classical oracle. An $\mathcal O$-oracle circuit is a quantum circuit that makes queries to $\mathcal O|_{\{0,1\}^{\leq n}}$ for some arbitrary $n$. An oracle circuit is a quantum circuit that makes queries to some unspecified oracle.

    A $\mathsf{P}$-uniform quantum oracle algorithm is defined as a collection of quantum oracle circuits $\{\mathcal A_\lambda\}_{\lambda\in \mathbb N_+}$ where the index $\lambda$ corresponds to the length of the input of the computational problem, such that there exists a deterministic polynomial time Turing machine $M$ that, on input $1^\lambda$, outputs the full classical description of the circuit $\mathcal A_\lambda$. In particular, $\mathcal A_\lambda$ has at most polynomially many oracle and elementary gates, and queries the oracle at length at most polynomial.
\end{definition}

We need to be careful on the difference between a (total) language and a promise language. Thus, we make the difference explicit by defining the \emph{promise} version of complexity classes first. Since a decision problem is naturally a promise problem (with no out-of-promise input), these definitions also extend to decision problems naturally.


\begin{definition}[Oracle $\qcma$]
    Given an oracle $\mathcal O$, a promise language $\mathcal L=(\mathcal L_{\text{YES}},\mathcal L_{\text{NO}})$ is in $\pqcma^{\mathcal O}$, if there exists a $\mathsf P$-uniform quantum oracle algorithm $\{\mathcal A_\lambda\}_{\lambda\in \mathbb N_+}$ accepting a classical witness of length $t(\lambda)$, such that for every input $x$ of length $\lambda=|x|$,
    \begin{itemize}
        \item $x\in \mathcal L_{\text{YES}}\implies \exists w\in \{0,1\}^{t(\lambda)},\ \Pr[\mathcal A_\lambda^{\mathcal O}(x,w)=1]\geq 2/3$.
        \item $x\in \mathcal L_{\text{NO}}\implies \forall w\in \{0,1\}^{t(\lambda)},\ \Pr[\mathcal A_\lambda^{\mathcal O}(x,w)=1]\leq 1/3$.
    \end{itemize}
    The class $\qcma^{\mathcal O}$ is defined as the subset of decision problems in $\pqcma^{\mathcal O}$.
\end{definition}


\begin{definition}[Oracle $\qma$]
    Given an oracle $\mathcal O$, a promise language $\mathcal L=(\mathcal L_{\text{YES}},\mathcal L_{\text{NO}})$ is in $\pqma^{\mathcal O}$, if there exists a $\mathsf P$-uniform quantum oracle algorithm $\{\mathcal A_\lambda\}_{\lambda\in \mathbb N_+}$ accepting a quantum witness of length $t(\lambda)$, such that for every input $x$ of length $\lambda=|x|$,
    \begin{itemize}
        \item $x\in \mathcal L_{\text{YES}}\implies \exists \ket{\phi}\in (\mathbb C^2)^{\otimes t(\lambda)},\ \Pr[\mathcal A_\lambda^{\mathcal O}(x,\ket{\phi})=1]\geq 2/3$.
        \item $x\in \mathcal L_{\text{NO}}\implies \forall \ket{\phi}\in (\mathbb C^2)^{\otimes t(\lambda)},\ \Pr[\mathcal A_\lambda^{\mathcal O}(x,\ket{\phi})=1]\leq 1/3$.
    \end{itemize}
    The class $\qma^{\mathcal O}$ is defined as the subset of decision problems in $\pqma^{\mathcal O}$.
\end{definition}

Next, we generalize $\mathsf{PH}$ to quantum--classical proof systems.
The most natural generalization is to define the $\qcma$ hierarchy as
$$\qcma\cup \qcma^\qcma \cup \qcma^{\qcma^\qcma}\cup\cdots\,.$$
However, the complete language for $\qcma$ is a promise problem, not a standard oracle (i.e.\ a total function). We handle this as follows.

\begin{definition}[Oracle $\qcmah$, following~\cite{AIK22,AB26}]\label{def:qcmah}
    Let $\mathcal O$ be an oracle. For every $i\in \mathbb N_+$, we inductively define the class $\pqcmah_i^{\mathcal O}$ as follows. The first level is defined as $$\pqcmah_1^{\mathcal O}:=\pqcma^{\mathcal O}\,.$$
    Next we define $\pqcmah_i^{\mathcal O}$. A promise language $\mathcal L=(\mathcal L_{\text{YES}},\mathcal L_{\text{NO}})$ is in $\pqcmah_i^{\mathcal O}$, if there exists a promise language $\Pi\in \pqcmah_{i-1}^{\mathcal O}$ and a $\mathsf P$-uniform quantum oracle algorithm $\{\mathcal A_\lambda\}_{\lambda\in \mathbb N_+}$ accepting a classical witness of length $t(\lambda)$, such that for every input $x$ of length $\lambda=|x|$,
    \begin{itemize}
        \item If $x\in \mathcal L_{\text{YES}}$: for any total completion $H$ of the promise language $\Pi$, 
        $$\exists w\in \{0,1\}^{t(\lambda)},\ \Pr[\mathcal A_\lambda^{H}(x,w)=1]\geq \frac23\,.$$
        \item If $x\in \mathcal L_{\text{NO}}$: for any total completion $H$ of the promise language $\Pi$,
        $$\forall w\in \{0,1\}^{t(\lambda)},\ \Pr[\mathcal A_\lambda^{H}(x,w)=1]\leq \frac13\,.$$
    \end{itemize}

    We define $$\pqcmah^{\mathcal O}:=\bigcup_{i=1}^\infty \pqcmah_i^{\mathcal O}\,,$$
    and the class $\qcmah^{\mathcal O}$ (resp.\ $\qcmah_i^{\mathcal O}$) is defined as the subset of decision problems in $\pqcmah^{\mathcal O}$ (resp.\ $\pqcmah_i^{\mathcal O}$).
\end{definition}

\begin{remark}\label{rmk:qcmah_alt}
    There are also different generalizations of $\mathsf{PH}$ to quantum--classical proof systems. For example, \cite{GSS+18} defines the class $\mathsf{QCPH}$ by constant layers of switching classical quantifiers. An alternative definition, denoted $\mathsf{QCPH'}$ in~\cite{AB26}, was formalized in~\cite{LG17}. Also, in our definition (Definition~\ref{def:qcmah}), the witness can depend on the specific total completion of the promise language. In particular, for a ``YES" instance, the $\qcma$ verifier may accept different proofs for different total completions. We could define a different class $\pqcmah_{\text{robust}}$ similar to Definition~\ref{def:qcmah} except that the ``YES" instance requires a single proof to work for all completions.

    It is not clear whether these classes are equal or not. Nevertheless,
    $$\mathsf{PromiseQCPH'}\subseteq \mathsf{PromiseQCPH}\subseteq\pqcmah_{\text{robust}}\subseteq \pqcmah\,.$$
    We focus on separating problems from the most powerful class $\qcmah$, and thus the result applies to all other variants.
\end{remark}

\subsection{Polynomials and sensitivity}

Following~\cite{AB26}, we define the fractional block sensitivity as follows. Note that we use a convention that differs from that of~\cite{AB26}.

\newcommand{\sen}{\operatorname{s}}
\newcommand{\fbs}{\operatorname{fbs}}

\begin{definition}\label{def:fbs}
    Given a real-valued function $f:\{0,1\}^n\to [0,1]$, for each $x\in \{0,1\}^n$ and $B\subseteq[n]$, define
    $$\sen(f,x,B)=\left|f(x)-f(x^B)\right|$$
    where
    $$x^B_i=\begin{cases}
        \overline{x_i} & i\in B\\
        x_i & i\notin B
    \end{cases}\,.$$

    Given a real-valued function $f:\{0,1\}^n\to [0,1]$, for each $x\in \{0,1\}^n$, define $\fbs(f,x)$ to be the solution of the following linear program:
    \begin{align*}
        \text{maximize}\quad &\sum_{B\subseteq[n]} w_B\sen(f,x,B) \\
        \text{subject to}\quad & w_B\geq 0 && \forall B\subseteq[n]\\
        & \sum_{B\ni i}w_B\leq 1 && \forall i\in [n]
    \end{align*}
    or equivalently, the following dual linear program:
    \begin{align*}
        \text{minimize} \quad & \sum_{i\in [n]} c_i\\
        \text{subject to}\quad & c_i\geq 0 && \forall i\in [n]\\
        & \sum_{i\in B}c_i\geq \sen(f,x,B) && \forall B\subseteq [n]
    \end{align*}
    Define the fractional block sensitivity (or fractional certificate complexity) to be
    $$\fbs(f):=\max_{x\in \{0,1\}^n} \fbs(f,x)\,.$$
\end{definition}

\begin{theorem}[\cite{AB26}]\label{thm:poly_bound_fbs}
    For any polynomial $f:\{0,1\}^n\to [0,1]$, we have
    $$\fbs(f)\leq \frac{\pi^2}{4}\deg(f)^2\,.$$
\end{theorem}

\subsection{Random restriction and H{\aa}stad switching lemma}

To formalize the H{\aa}stad switching lemma, we introduce the following definitions.

\begin{definition}[DNF and CNF]
A \emph{literal} over variables $x_1,x_2,\cdots,x_n$ is either $x_i$
or $\neg x_i$ for some $i\in[n]$.

A Boolean formula $F$ over variables $x=(x_1,x_2,\cdots,x_n)$ is in \emph{disjunctive normal form} (DNF) if
it is a disjunction of conjunctions of literals:
\[
F(x)=\bigvee_{j=1}^{m}\ \bigwedge_{\ell\in T_j}\ell(x).
\]
The DNF has
\emph{width} at most $w$ if $|T_j|\le w$ for every $j$.

A Boolean formula $G$ over variables $x=(x_1,x_2,\cdots,x_n)$ is in \emph{conjunctive normal form} (CNF) if
it is a conjunction of disjunctions of literals:
\[
G(x)=\bigwedge_{j=1}^{m}\ \bigvee_{\ell\in C_j}\ell(x).
\]
The CNF has
\emph{width} at most $w$ if $|C_j|\le w$ for every $j$.
\end{definition}

\begin{definition}
    Given a function $f:\{0,1\}^N\to R$, a restriction $\rho$ is any function from $[N]$ to $\{0,1,*\}$. Let $D=\rho^{-1}(*)$. The restricted function $f_\rho:\{0,1\}^{D}\to R$ is defined as
    $$f_\rho(x)=f(x_\rho)$$
    where
    $$(x_\rho)_i=\begin{cases}
        x_i & i\in D\\
        \rho(i) & i\notin D
    \end{cases}\,.$$
\end{definition}

\begin{definition}
    Let $f:\{0,1\}^k\to \{0,1,\perp\}$ be a partial function. Define
    $$\dt(f)=\min\{\operatorname{depth}(\mathcal T):\forall x\in f^{-1}(\{0,1\}),\ \mathcal T(x)=f(x)\}\,,$$
    where $\mathcal T$ is a deterministic decision tree.
\end{definition}

\begin{definition}
    Let $f:\{0,1\}^k\to [0,1]$ be a real-valued function. For any $\eta>0$, define
    $$\dt_\eta(f)=\min\{\operatorname{depth}(\mathcal T):\forall x\in \{0,1\}^k,\ |\mathcal T(x)-f(x)|<\eta\}\,,$$
    where $\mathcal T$ is a deterministic decision tree with real-valued leaves.
\end{definition}

We now state the H{\aa}stad switching lemma, slightly generalized by~\cite{Kelley21}. This generalization would be helpful in our proof of separating $\qma$ from $\qcmah$.

\begin{definition}[Adapted from~\cite{Kelley21}, Definition~1.5]\label{def:boundedness}
    Let $N\in \mathbb N_+$, and let $\mathcal D$ be a distribution over subsets of $[N]$. We say $\mathcal D$ is $t$-wise $p$-bounded, if
    $$\forall S\subseteq [N],\ |S|\leq t,\quad \Pr_{D\sim \mathcal D}[S\subseteq D]\leq p^{|S|}\,.$$
\end{definition}

\begin{theorem}[\cite{Kelley21}, Theorem~1.6]\label{thm:kelley-switching}
    Let $F:\{0,1\}^N\to \{0,1\}$ be a DNF (or CNF) over $N$ variables of width at most $w$. Fix $t\in \mathbb N_+$. Let $\mathcal D$ be a $(w+t)$-wise $p$-bounded distribution over subsets of $[N]$. For any $x\in \{0,1\}^N$ and $D\subseteq [N]$, define the restriction $\rho=\rho_{x,D}$ as
    $$\rho(i)=\begin{cases}
        * & i\in D\\
        x_i & i\notin D
    \end{cases}\,.$$
    Then,
    $$\Pr_{\substack{x\gets\{0,1\}^N\\ D\sim \mathcal D}}\left[\dt(F_\rho)\geq t\right]\leq (16pw)^t\,.$$
\end{theorem}


\subsection{Coding theory}

\begin{definition}[List recoverability]
    A code $C\subseteq \Sigma^n$ is $(\ell,L)$-list recoverable, if for any subsets $S_1,S_2,\cdots,S_n\subseteq \Sigma$,
    $$\frac{1}{n}\sum_{i=1}^n|S_i|\leq \ell \quad \implies \quad \left|C\cap (S_1\times S_2\times \cdots \times S_n)\right|\leq L\,.$$
\end{definition}

\begin{theorem}[\cite{BHV26}, Corollary~4.18]\label{thm:list-recoverable-mult}
    For each parameter $\lambda\in \mathbb N_+$, let $k=\lambda^3$, $\lambda^5<q\leq 2\lambda^5$ be any prime, and $s=\lambda$. Then, for sufficiently large $\lambda$, $C_\lambda=\mult_{s,\mathbb F_q,k}$ is $(\ell,2\ell)$-list recoverable for all $\ell\leq 2^\lambda$, where $\mathsf{Mult}_{s,\mathbb F_q,k}$ is the multiplicity code defined in Definition~\ref{def:mult-code}.
\end{theorem}

\subsection{Concentration theorems}

\begin{theorem}[Matrix Bernstein inequality, Theorem~6.1.1 in~\cite{Tropp-matrix-concentration}]\label{thm:matrix-bernstein}
    Let $\{X_k\}$ be a sequence of independent random Hermitian matrices of dimension $d$. Assume for each index $k$,
    $$\EE X_k=0\,,\quad \|X_k\|\leq L\,.$$
    Let
    $$Y=\sum_k X_k\,,$$
    and define
    $$\Var(Y)=\|\EE Y^2\|=\left\|\sum_k \EE X_k^2\right\|\,.$$
    Then,
    $$\Pr[ \|Y\|\geq \eta]\leq 2d\exp\left(\frac{-\eta^2/2}{\Var(Y)+L\eta/3}\right)\,.$$
\end{theorem}

%% file: Proof/quantum-switching.tex
In this section, we present the polynomial switching lemma (Theorem~\ref{thm:quantum_switching}) and its applications to $\mathsf{BQP}$ and $\qcmah$ algorithms.

\subsection{The polynomial switching lemma}
\begin{theorem}\label{thm:quantum_switching}
    Let $f:\{0,1\}^N\to [0,1]$ be a real-valued function. Fix any $x\in \{0,1\}^N$ and $t\in \mathbb N_+$. Let $\mathcal D$ be a $t$-wise $p$-bounded distribution over subsets of $[N]$ (see Definition~\ref{def:boundedness}), which means
    \begin{equation}\label{eq:spread_of_D}
        \forall S\subseteq[N],\ |S|\leq t,\quad \Pr_{D\sim \mathcal D}[S\subseteq D]\leq p^{|S|}\,.
    \end{equation}
    Define the restriction $\rho=\rho_{x,D}$ as
    $$\rho(i)=\begin{cases}
        * & i\in D\\
        x_i & i\notin D
    \end{cases}\,.$$
    Then, for all $\eta>0$,
    $$\Pr_{D\sim \mathcal D}\left[\dt_\eta(f_\rho)\geq t\right]\leq \left(\frac{2p\fbs(f)}{\eta}\right)^t\,.$$
\end{theorem}

\begin{proof}
    Fix \(x,t\). We construct a sampler $\mathsf{Samp}$ that, whenever the ``bad'' event \(\mathrm{DT}_\eta(f_\rho)\ge t\) occurs, outputs \(t\) distinct points in \(D\) with abnormally high probability. We then use \eqref{eq:spread_of_D} to bound the probability of this ``bad'' event.

    By the definition of fractional certificate complexity (Definition~\ref{def:fbs}), for all $x$ and $i$, there exists $c_{i,x}\geq 0$ such that for all $y$,
    \begin{equation}\label{eq:fbs_definition}
        |f(x)-f(y)|\leq\sum_{i:x_i\neq y_i}c_{i,x}\,,\quad \sum_{i\in [N]}c_{i,x}\leq \fbs(f)\,.
    \end{equation}
    Note that $c_{i,x}$ only depends on the function $f$.
    
    The sampler $\mathsf{Samp}$ is defined as follows.
    \begin{algorithm}[H]
        \caption{The sampler $\mathsf{Samp}$}
        \label{alg:sampler}
        \begin{algorithmic}
            \State \textbf{Input:} $f,x,t$
            \State $x_0:=x$
            \State $S_0:=\varnothing$
            \For{$k=1,2,\cdots,t$}
                \State Sample $i_k$ with probability $\Pr[i]=c_{i,x_{k-1}}/\fbs(f)$
                \State $S_k:=S_{k-1}\cup \{i_k\}$
                \State Let $x_k:=x_{k-1}$ except that $(x_k)_{i_k}\gets \{0,1\}$ uniformly
            \EndFor
            \State \textbf{Output} $S_t$
        \end{algorithmic}
    \end{algorithm}
    
    Note that in Algorithm~\ref{alg:sampler}, the probability distribution of each $i_k$ is valid due to~\eqref{eq:fbs_definition}. Also, Algorithm~\ref{alg:sampler} may be inefficient, but it does not get any information about $D$ from its input, which will be the information-theoretic property that the proof relies on.

    We claim that
    \begin{equation}\label{eq:samp_succeeding_prob}
        \Pr[|\mathsf{Samp}(f,x,t)|=t\land\mathsf{Samp}(f,x,t)\subseteq D\mid \dt_\eta(f_\rho)\geq t]\geq \left(\frac{\eta}{2\fbs(f)}\right)^t\,.
    \end{equation}

    To prove~\eqref{eq:samp_succeeding_prob}, we fix $D$ such that $\dt_\eta(f_\rho)\geq t$. Let $\rho_0=\rho$; let $\rho_k$ be the same as $\rho_{k-1}$, except that its $i_k$-th entry is replaced by the value of $(x_k)_{i_k}$ sampled from $\{0,1\}$. Define the event $P_0:=\top$, and
    $$P_k:=P_{k-1}\land (i_k\in D\backslash S_{k-1}) \land (\dt_\eta(f_{\rho_k})\geq t-k)\,.$$
    It suffices to prove
    $$\Pr[P_k\mid P_{k-1}]\geq \frac{\eta}{2\fbs(f)}\,.$$
    Suppose from now on that $P_{k-1}$ is satisfied. This means $\dt_\eta(f_{\rho_{k-1}})\geq t-k+1$. In particular, $\dt_\eta(f_{\rho_{k-1}})\geq1$. This means that there exist $y_0,y_1$ satisfying the random restriction $\rho_{k-1}$, while
    $$|f(y_0)-f(y_1)|\geq 2\eta\,.$$
    Thus, there exists $y$ such that
    $$|f(y)-f(x_{k-1})|\geq \eta$$
    while
    $$\{i:(x_{k-1})_i\neq y_i\}\subseteq \rho_{k-1}^{-1}(*)=D\backslash S_{k-1}\,.$$
    By~\eqref{eq:fbs_definition},
    $$\sum_{i\in D\backslash S_{k-1}}c_{i,x_{k-1}}\geq \sum_{i:(x_{k-1})_i\neq y_i}c_{i,x_{k-1}}\geq \eta\,.$$
    And hence, the probability that $i_k\in D\backslash S_{k-1}$ is greater than $\eta/\fbs(f)$.

    Now we show that $\dt_\eta(f_{\rho_k})\geq t-k$ with probability at least $1/2$, if $i_k\in D\backslash S_{k-1}$. But this is simply because 
    $$\dt_\eta(g)\leq 1+\max\left(\dt_\eta(g|_{\{x=0\}}),\dt_\eta(g|_{\{x=1\}})\right)$$
    for any function $g$ and any $x$. This concludes the proof of~\eqref{eq:samp_succeeding_prob}.

    Notice that, by~\eqref{eq:spread_of_D},
    $$\Pr[|\mathsf{Samp}(f,x,t)|=t\land\mathsf{Samp}(f,x,t)\subseteq D]\leq p^t\,.$$
    Combining this bound with~\eqref{eq:samp_succeeding_prob} yields the desired theorem.
\end{proof}

\begin{corollary}\label{cor:quantum_switching}
    Let $L:\{0,1\}^N\to \{0,1,\perp\}$ be a partial function which has a bounded-error $T$-query quantum algorithm. Fix any $x\in \{0,1\}^N$ and $t\in \mathbb N_+$. Let $\mathcal D$ be a $t$-wise $p$-bounded distribution over subsets of $[N]$.
    Define the restriction $\rho=\rho_{x,D}$ as
    $$\rho(i)=\begin{cases}
        * & i\in D\\
        x_i & i\notin D
    \end{cases}\,.$$
    Then
    $$\Pr_{D\sim \mathcal D}[\dt(L_\rho)\geq t]\leq \left(12\pi^2pT^2\right)^t\,.$$
\end{corollary}
\begin{proof}
    The acceptance probability of the $T$-query quantum algorithm can be represented by a polynomial $f$ of degree at most $2T$. By Theorem~\ref{thm:poly_bound_fbs},
    $$\fbs(f)\leq \pi^2T^2\,.$$
    And a decision tree with $\eta=1/6$ suffices to determine exactly whether the value of $f$ is greater than $2/3$ or smaller than $1/3$. In other words,
    $$\dt(L_\rho)\leq \dt_{1/6}(f_\rho)\,.$$
    Theorem~\ref{thm:quantum_switching} yields the desired bound.
\end{proof}

\subsection{Random restriction for $\qcmah$}\label{sec:switching-qcmah}

We recall the standard result~\cite{FSS84} in classical complexity theory that any oracle $\mathsf{PH}$ algorithm can be interpreted as a quasipolynomially large $\mathsf{AC}^0$ circuit. Similarly, we have the following result for $\mathsf{QCMAH}$, clarifying the formulation in~\cite[Proposition~6.6]{AB26}.

We use the following convention when a circuit contains partial functions.

\begin{definition}\label{def:partial-func-extension}
Let $\mathcal C$ be some circuit whose gates may compute some partial functions.
A \emph{consistent completion} $\widehat {\mathcal C}$ is obtained by replacing each gate by a total completion of the function computed by this gate.

For every input $x$,
define \[
\mathcal C(x)=
\begin{cases}
0 & \text{if every consistent completion }
    \widehat{\mathcal C}\text{ satisfies }\widehat{\mathcal C}(x)=0,\\
1 & \text{if every consistent completion }
    \widehat{\mathcal C}\text{ satisfies }\widehat{\mathcal C}(x)=1,\\
\perp & \text{otherwise}.
\end{cases}
\]
\end{definition}

\begin{proposition}\label{prop:qcmah-circuits}
    Let $\mathcal L:\{0,1\}^*\to \{0,1,\perp\}$ be a promise problem in $\pqcmah^{\mathcal O}_d$. Then, there exists a polynomial $q$ such that the following holds. Let $x\in \{0,1\}^n$ be a fixed input such that $\mathcal L(x)\neq{} \perp$. Then, there exists a circuit $\mathcal C$ of size $2^{q(n)}$ and depth $2d$, independent of $\mathcal O$, satisfying the following:
    \begin{itemize}
        \item It takes the truth table of $\mathcal O_{\leq q(n)}$ as input.
        \item Each gate is either an OR gate (of unbounded fan-in), or a \emph{partial} function that can be computed by a $q(n)$-query quantum algorithm.
        \item $\mathcal C(\mathcal O_{\leq q(n)})=\mathcal L(x)$, where the evaluation of $\mathcal C$ follows Definition~\ref{def:partial-func-extension}; that is, every \emph{consistent} completion of $\mathcal C$ evaluates to $\mathcal L(x)$ on input $\mathcal O_{\leq q(n)}$.
    \end{itemize}
\end{proposition}
\begin{proof}
    We prove the proposition by induction on $d$.
    
    The base case is $d=1$, and $\pqcmah_d^{\mathcal O}=\pqcma^{\mathcal O}$. This means $\mathcal L$ can be decided by a $\qcma$ verifier $V$, which accepts a classical witness of length at most $p(n)$, and runs a quantum circuit with at most $p(n)$ queries to $\mathcal O_{\leq p(n)}$, for some polynomial $p$. This can be interpreted as taking the OR over all $2^{p(n)}$ witnesses, with each choice of witness corresponding to a partial function computed by $V^{(\cdot)}(x,w)$.

    Now, suppose that for any promise problem $\Pi\in \pqcmah_{d-1}^{\mathcal O}$, there exists a polynomial $q$ that satisfies the proposition. Then, $\mathcal L$ can be decided by a $\mathsf{QCMA}$ verifier $V$, which accepts a classical witness of length at most $p(n)$, and runs a quantum circuit with at most $p(n)$ queries to $H_{\leq p(n)}$ for some polynomial $p$, where $H$ is a decision problem that completes $\Pi$. We convert them into circuits as follows:
    \begin{itemize}
        \item We convert $V$ into an OR of partial functions, similar to the base case (note that the circuit is independent of the total completion $H$).
        \item For each string $z$ queried by $V$, using the inductive hypothesis, we convert $\Pi(z)$ into a circuit of size $2^{q(|z|)}$ and depth $2(d-1)$, querying $\mathcal O_{\leq q(|z|)}$, where each gate is either an OR gate or a partial function that can be computed by a $q(|z|)$-query quantum algorithm, and $q$ is some polynomial.
    \end{itemize}
    Definition~\ref{def:partial-func-extension} turns out to be exactly the consistency condition in Definition~\ref{def:qcmah}. In particular,
    \begin{itemize}
        \item the circuit evaluates to $1$ if for every total completion $H$ of $\Pi$, there exists a witness $w$ (that may depend on the total completion) such that $V^{H}(x,w)$ accepts with probability $\geq2/3$;
        \item the circuit evaluates to $0$ if for every total completion $H$ of $\Pi$ and for any witness $w$, $V^{H}(x,w)$ accepts with probability $\leq 1/3$.
    \end{itemize}
    Since $q(p(n))$ is still a polynomial in $n$, this proves the proposition for $\pqcmah_d$.
\end{proof}

Now we prove that any circuit of the form described in Proposition~\ref{prop:qcmah-circuits} can be reduced to a constant function after a series of random restrictions.

\begin{theorem}\label{thm:qcmah-restriction}
    Suppose $F:\{0,1\}^N\to \{0,1,\perp\}$ can be computed by a circuit of size $s$ and depth $d$, where each gate is either an OR gate (of unbounded fan-in), or a \emph{partial} function that can be computed by a $T$-query quantum algorithm.
    
    Let $t\in \mathbb N_+$. Suppose there exists a distribution of families of sets $D_0\supseteq D_1\supseteq\cdots\supseteq D_{d+1}$ such that for all $k\in [d+1]$, the distribution of each $D_k$ is $2t$-wise $p$-bounded, conditioned on $D_{k-1}$. Let $x$ be sampled uniformly from $\{0,1\}^N$, and define the restriction $\rho_{k}=\rho_{x,D_{k}}$ as
    $$\rho_k(i)=\begin{cases}
        * & i\in D_k\\
        x_i & i\notin D_k
    \end{cases}\,.$$
    Then
    \begin{equation}\label{eq:qcmah-restriction-1}
        \Pr[\dt(F_{\rho_d})\geq t]\leq s(1000pT^2t^2)^t\,.
    \end{equation}
    Furthermore,
    \begin{equation}\label{eq:qcmah-restriction-2}
        \Pr[\dt(F_{\rho_{d+1}})\geq 1]\leq s(1000pT^2t^2)^t+p(t-1)\,.
    \end{equation}
    In other words, with high probability over $x$ and $D_{d+1}$, the restriction of $F$ on $\rho_{d+1}$ is constant on its promised inputs.
\end{theorem}

\begin{proof}
    We first prove~\eqref{eq:qcmah-restriction-1} by induction.

    Define the \emph{depth} of a gate as follows. Each input bit is at depth $0$. A gate has depth at most $k$, if it can be computed from gates at depth not greater than $k-1$. Let $s_k$ be the total number of gates of depth not greater than $k$ in the circuit computing $F$. The inductive hypothesis is as follows.

    \begin{itemize}
        \item \textbf{Inductive hypothesis.} With probability $\geq1-s_k(1000pT^2t^2)^t$, the following holds. For all gates $g$ of depth not greater than $k$, let $G$ denote the subcircuit computing the value of $g$. Then,
        \begin{itemize}
            \item There exists a decision tree family $\mathcal A=\{A_g\}$, indexed by gates $g$ of depth not greater than $k$, such that the depth of $A_g$ is less than $t$, and $A_g$ computes $G_{\rho_k}$;
            \item Moreover, for each input $y$ in the restricted subcube of $\rho_k$, there exists a consistent completion (Definition~\ref{def:partial-func-extension}) of the circuit (which may depend on the input), such that for all $g$, $A_g(y)$ equals the value of $g$ under that consistent completion.
        \end{itemize}
    \end{itemize}
    The base case $k=0$ is trivial. Also, it is clear that \eqref{eq:qcmah-restriction-1} is implied by the inductive hypothesis when $k=d$.
    
    From now on, fix \(k<d\) and assume the inductive hypothesis holds for \(k\). We prove the inductive hypothesis for \(k+1\).
    
    We choose $\rho_k$ satisfying the condition in the inductive hypothesis. This holds with probability at least $1-s_k(1000pT^2t^2)^t$.
    We fix the decision tree family $\mathcal A$ defined in the inductive hypothesis, and replace each gate of depth $\leq k$ by its corresponding decision tree in $\mathcal A$.
    
    Let $g$ be a gate of depth $k+1$, and $G$ be the corresponding subcircuit.

    \begin{itemize}
        \item \textbf{The case when $g$ is an OR gate.} By assumption, under the restriction $\rho_k$, this gate can be computed by an OR over depth-$t$ decision trees. In other words, there is a width-$t$ DNF computing the value of this gate on $\rho_k$. Since the distribution of $D_{k+1}$ is $2t$-wise $p$-bounded in $D_k$, by Theorem~\ref{thm:kelley-switching},
        $$\Pr[\dt(G_{\rho_{k+1}})\geq t]\leq (16pt)^t\,.$$
        \item \textbf{The case when $g$ is a $T$-query quantum algorithm.} By assumption, under the restriction $\rho_k$, this gate can be computed by a quantum algorithm that makes $T$ queries to some depth-$t$ decision trees (that is, $t$-query classical algorithms). This yields a $2Tt$-query\footnote{The factor of $2$ comes from the cost of uncomputing the work register when quantumly simulating the decision trees.} quantum algorithm computing the same gate. By Corollary~\ref{cor:quantum_switching},
        $$\Pr[\dt(G_{\rho_{k+1}})\geq t]\leq (12\pi^2p(2Tt)^2)^t\,.$$
    \end{itemize}
    Thus, in summary, each subcircuit $G$ can be reduced to a decision tree of depth less than $t$ when restricting to $\rho_{k+1}$, except with probability at most $(1000pT^2t^2)^t$. Moreover, for every input $y$ consistent with $\rho_{k+1}$, there exists a total completion of $g$ such that $G(y)$ is evaluated by the decision tree.\footnote{A total completion of a subcircuit does not induce a single total completion of each gate that works for all inputs. However, the inductive hypothesis is consistent with Definition~\ref{def:partial-func-extension}, which automatically means that a consistent completion can be found for each individual input. This is why we require the following order of quantifiers in the inductive hypothesis: for each fixed $\rho_k$, there exist a decision tree family $\{A_g\}$ such that, for all input $y$, there exists a consistent completion.} Therefore, the inductive hypothesis holds also for depth $k+1$ using a union bound.

    Now that we have proved~\eqref{eq:qcmah-restriction-1}, we prove~\eqref{eq:qcmah-restriction-2} as follows. Fix $x$ and assume $\dt(F_{\rho_d})<t$. Let $A$ be a decision tree of depth less than $t$ for $F_{\rho_d}$. Since $x$ is already fixed, the query path for $A(x_{\rho_d})$ is also fixed. In particular, there exist coordinates $I=\{i_1,i_2,\cdots, i_{t-1}\}$, such that for every $y$, $A$ returns the same value once $y|_I=x|_I$. And therefore, conditioned on $D_d$ and $x$ such that $\dt(F_{\rho_d})<t$,
    $$\Pr[\dt(F_{\rho_{d+1}})=0]\geq \Pr[I\subseteq D_d\backslash D_{d+1}]\,.$$
    Since $D_{d+1}$ is $1$-wise $p$-bounded, each coordinate is in $D_{d+1}$ with probability at most $p$, and thus \eqref{eq:qcmah-restriction-2} follows by a union bound.
\end{proof}

\begin{remark}\label{rmk:conventions}
    The proof will be simpler if we define the evaluation of a circuit with partial functions as in~\cite[Section~2.5]{AIK22}, which is different from Definition~\ref{def:partial-func-extension}. However, consider the same circuit constructed in Proposition~\ref{prop:qcmah-circuits}. The \cite{AIK22} convention corresponds to $\pqcmah_{\text{robust}}$ (as defined in Remark~\ref{rmk:qcmah_alt}) instead of $\pqcmah$. As we will see in Section~\ref{sec:collapse}, to prove Theorem~\ref{thm:collapse}, it is crucial to use the consistent completion definition (Definition~\ref{def:partial-func-extension}).
\end{remark}

%% file: Proof/proof-main.tex
\paragraph{Separating oracle.} We use, nearly as a black box, the code intersection problem in~\cite{BHV26}. For each parameter $\lambda\in \mathbb N_+$, let $C_\lambda=\mult_{s,\mathbb F_q,k}$ be the multiplicity code defined as in Definition~\ref{def:mult-code}, where $k=\lambda^3$, $q$ is the smallest prime greater than $\lambda^5$, and $s=\lambda$. In particular, $\lambda^5<q\leq 2\lambda^5$. Define $\Sigma=\mathbb F_q^s$.

We embed $\{0,1\}^\lambda\times \Sigma^q$ into $\{0,1\}^{\lambda^7}$, as discussed in~\cite[Remark~6.2]{BHV26}.

\begin{definition}[Separating problem]\label{def:separating-problem}
For any function $H:[q]\times\Sigma\to \{0,1\}$, define
$$D_H=\left\{\begin{aligned}
    (x,v):{}&v\in C_\lambda\,,\\
    &\forall i\in \{1,\cdots,\lambda\},\ H(i,v_i)=x_i\,,\\
    &\forall i\in \{\lambda+1,\cdots,q\},\ H(i,v_i)=0
\end{aligned}\right\}\subseteq\{0,1\}^{\lambda}\times \Sigma^q\subseteq \{0,1\}^{\lambda^7}\,.$$

We define the following distributions over functions $f_\lambda:\{0,1\}^{\lambda^7}\to \{0,1\}$.
\begin{itemize}
    \item (``YES" distribution $\mathscr D_{\text{YES}}$) $f_\lambda=R\lor \mathbf{1}_{D_H}$, where $R:\{0,1\}^{\lambda^7}\to \{0,1\}$ is an independently uniformly sampled function, and $H:[q]\times\Sigma\to \{0,1\}$ is sampled with each bit equal to $1$ independently with probability $p=\lambda^{-4}$.
    \item (``NO" distribution $\mathscr D_{\text{NO}}$) $f_\lambda=R$, where $R:\{0,1\}^{\lambda^7}\to \{0,1\}$ is an independently uniformly sampled function.
\end{itemize}
\end{definition}

We will see that a $\mathsf{QMA}$ algorithm distinguishes ``YES" distribution and ``NO" distribution, but no $\mathsf{QCMAH}$ algorithm can.

\subsection{The $\qma$ algorithm}\label{sec:proof-main-qma}
To distinguish the ``YES" distribution and the ``NO" distribution, we use the biased Yamakawa--Zhandry algorithm $\byz$ exactly as in~\cite{BHV26}. The definition can be found in Algorithm~\ref{alg:biasedYZ}.

We use the following lemma, which shows that Algorithm~\ref{alg:biasedYZ} outputs points in $D_H$ with high probability, if the advice state is chosen properly.

\begin{lemma}[\cite{BHV26}, Corollary~5.24]\label{lem:Biased-YZ-succeeding}
    The following holds for sufficiently large $\lambda$. With probability $1-2^{-\lambda}$ over the choice of $H$, there exists $\ket{\adv_H}$ such that for any $x\in \{0,1\}^\lambda$,
    $$ \Pr_{v\gets\byz(\ket{\mathsf{adv}_H},x)}\left[\left(x, v\right)\in D_H\right]\geq 1-2^{-\lambda}\,.$$
\end{lemma}

Now we construct a $\mathsf{QMA}$ distinguisher of the ``YES" distribution and the ``NO" distribution.

\begin{algorithm}[H]
    \caption{The $\mathsf {QMA}$ verifier $V^{f_\lambda}$}
    \label{alg:QMA-ver}
    \begin{algorithmic}
        \State \textbf{Input:} The parameter $1^\lambda$, a quantum witness $\ket{\adv}$
        \State Uniformly sample $x\gets \{0,1\}^\lambda$
        \State Run $v\gets \byz(\ket{\adv},x)$
        \State \textbf{Output} $f_\lambda(x,v)$
    \end{algorithmic}
\end{algorithm}

\begin{corollary}\label{cor:qma-yes}
    For sufficiently large $\lambda$,
    $$\Pr_{f_\lambda\gets \mathscr D_{\text{YES}}}\left[\max_{ \ket{\mathsf{adv}}} \Pr\left[V^{f_\lambda}(1^\lambda,\ket{\mathsf{adv}})=1\right]\geq 1-2^{-\lambda}\right]\geq 1-2^{-\lambda}\,.$$
\end{corollary}
\begin{proof}
    This is a direct corollary of Lemma~\ref{lem:Biased-YZ-succeeding}.
\end{proof}

\begin{lemma}\label{lem:qma-no}
    Fix $0<\eta<1/2$. For sufficiently large $\lambda$,
    $$\Pr_{f_\lambda\gets \mathscr D_{\text{NO}}}\left[\max_{ \ket{\mathsf{adv}}} \Pr\left[V^{f_\lambda}(1^\lambda,\ket{\mathsf{adv}})=1\right]\geq \frac{1}{2}+\eta\right]\leq\exp\left(-\eta^2\cdot 2^\lambda\right)\,.$$
\end{lemma}
\begin{proof}
    We think of the biased Yamakawa--Zhandry algorithm $\byz$ abstractly as follows: the algorithm takes $x$ and a state $\ket{\adv}$ of $w$ qubits, performs a POVM measurement $\{E_{x,v}\}_{v\in \Sigma^q}$ on the state $\ket{\adv}$, and outputs the measurement result. Then, for a uniformly random $R:\{0,1\}^{\lambda^7}\to \{0,1\}$,
    $$\Pr\left[V^{R}(1^\lambda,\ket{\adv})=1\right]=\Tr\left[\frac{1}{2^\lambda}\sum_{x\in \{0,1\}^\lambda}\sum_{v\in \Sigma^q}R(x,v)E_{x,v}\ketbra{\adv}{\adv}\right]\,,$$
    and thus
    $$\max_{\ket{\adv}}\Pr\left[V^{R}(1^\lambda,\ket{\adv})=1\right]=\left\|\frac{1}{2^\lambda}\sum_{x\in \{0,1\}^\lambda}\sum_{v\in \Sigma^q}R(x,v)E_{x,v}\right\|\,.$$
    We bound the spectral norm of the right hand side using the matrix Bernstein inequality. Let
    $$X_{x,v}=\frac{1}{2^\lambda}\left(R(x,v)-\frac{1}{2}\right)E_{x,v}\,,$$
    and we have that for all $x,v$,
    $$\EE X_{x,v}=0\,,\quad \|X_{x,v}\|\leq \frac{1}{2^{\lambda+1}}\,.$$
    Furthermore,
    \begin{align*}
        \sum_{x,v}X_{x,v}^2={}&\sum_{x,v}\frac{1}{2^{2\lambda}}\left(R(x,v)-\frac{1}{2}\right)^2E_{x,v}^2\\
        \preceq {}& \sum_{x,v}\frac{1}{2^{2\lambda+2}}E_{x,v}^2\\
        \preceq {}& \sum_{x,v}\frac{1}{2^{2\lambda+2}}E_{x,v}\\
        ={}& \frac{1}{2^{2\lambda+2}}\sum_{x\in \{0,1\}^\lambda}\sum_{v\in \Sigma^q}E_{x,v}\\
        \preceq{}& \frac{I}{2^{\lambda+2}}\,.
    \end{align*}
    Therefore, by the matrix Bernstein inequality (Theorem~\ref{thm:matrix-bernstein}),
    $$\Pr\left[\left\|\sum_{x\in \{0,1\}^\lambda}\sum_{v\in \Sigma^q}X_{x,v}\right\|\geq \eta\right]\leq 2^{w+1}\cdot \exp\left(-\frac{3}{2}\eta^2\cdot 2^\lambda\right)\,.$$
    In particular, the length $w$ of the $\byz$ advice state is a fixed polynomial in $\lambda$, and thus for sufficiently large $\lambda$,
    $$\Pr\left[\left\|\frac{I}{2}-\frac{1}{2^\lambda}\sum_{x\in \{0,1\}^\lambda}\sum_{v\in \Sigma^q}R(x,v)E_{x,v}\right\|\geq \eta\right]\leq  \exp\left(-\eta^2\cdot 2^\lambda\right)\,.$$
    And the lemma follows by a triangle inequality.
\end{proof}

\subsection{The circuit lower bound}

Fix any $d\in \mathbb N_+$. We sample the sets $D_{d,0},D_{d,1},\cdots,D_{d,d}$ as follows.

We first sample $H_{\text{head}}:[\lambda]\times \Sigma\to \{0,1\}$, where each bit equals $1$ independently with probability $p=\lambda^{-4}$. Define
$$D_{d,0}=\left\{\begin{aligned}
    (x,v):{}&v\in C_\lambda\,,\\
    &\forall i\in \{1,\cdots,\lambda\},\ H_{\text{head}}(i,v_i)=x_i
\end{aligned}\right\}\,.$$
Next, sample $T_k:\{\lambda+1,\cdots,q\}\times \Sigma\to \{0,1\}$ for $k\in [d]$, where each bit equals $1$ independently with probability $1-(1-p)^{1/d}$. Define
$$H_{\text{tail}}^{(k)}=\bigvee_{j=1}^k T_j\,,$$
$$D_{d,k}=\left\{\begin{aligned}
    (x,v):{}&v\in C_\lambda\,,\\
    &\forall i\in \{1,\cdots,\lambda\},\ H_{\text{head}}(i,v_i)=x_i\,,\\
    &\forall i\in \{\lambda+1,\cdots,q\},\ H_{\text{tail}}^{(k)}(i,v_i)=0
\end{aligned}\right\}\,.$$

For any $d$, clearly $D_{d,0}\supseteq D_{d,1}\supseteq\cdots\supseteq D_{d,d}$, and for a random $R$, the function $R\lor \mathbf{1}_{D_{d,d}}$ has the same distribution as the ``YES" distribution of the separating problem.

We prove the following lemma, asserting that a fixed set of points in \(D_{d,k-1}\) is unlikely to be contained in \(D_{d,k}\).

\begin{lemma}\label{lem:sampling_D_probability}
    The following holds for sufficiently large $\lambda$. Fix $d\in \mathbb N_+$ and $k\in [d]$. For all $S\subseteq \{0,1\}^{\lambda^7}$ with $|S|\leq 2^\lambda$, conditioned on any choice of $D_{d,k-1}$,
    $$\Pr[S\subseteq D_{d,k}\mid S\subseteq D_{d,k-1}]\leq \exp\left(-\frac{\lambda}{3d}|S|\right)\,.$$
    In particular, conditioned on $D_{d,k-1}$, the distribution of $D_{d,k}$ is $2^\lambda$-wise $e^{-\lambda/3d}$-bounded.
\end{lemma}
\begin{proof}
    We observe that
    $$D_{d,k}=\{(x,v)\in D_{d,k-1}:\forall i\in \{\lambda+1,\cdots,q\},\,T_k(i,v_i)=0\}\,.$$
    Fix $H_{\text{head}}$ and $T_1,\cdots,T_{k-1}$. Choose $S\subseteq D_{d,k-1}$ with $|S|\leq 2^\lambda$. For $i\in [q]$, define
    $$S^*_i=\{v_i:\exists x,(x,v)\in S\}\,.$$
    We have
    \begin{align*}
        \Pr[S\subseteq D_{d,k}]={}&\prod_{i=\lambda+1}^q \Pr[T_k(i,S_i^*)=0]\\
        ={}& \left((1-p)^{1/d}\right)^{\sum_{i=\lambda+1}^q |S_i^*|}\,.
    \end{align*}
    Note that $x$ is determined by $v$ once $H_{\text{head}}$ is fixed. Thus,
    $$|S_i^*|\leq |S|\,.$$
    By the list recoverability of $C_\lambda$ (Theorem~\ref{thm:list-recoverable-mult}),
    $$\sum_{i=1}^q |S_i^*|\geq \frac{q}{2}|C_\lambda\cap (S_1^*\times S_2^*\times\cdots \times S_q^*)|\geq \frac{q|S|}{2}\,,$$
    and thus
    $$\sum_{i=\lambda+1}^q |S_i^*|\geq \left(\frac{q}{2}-\lambda\right)|S|\geq \frac{\lambda^5}{3}|S|\,.$$
    Therefore,
    \begin{align*}
        \Pr[S\subseteq D_{d,k}]\leq{}& \left((1-\lambda^{-4})^{\lambda^5/3d}\right)^{|S|}\\
        \leq{}& \exp\left(-\frac{\lambda}{3d}|S|\right)\,.
    \end{align*}
\end{proof}

Now we apply the iterative restriction defined by $D_{d,k}$ to bound the probability of a $\mathsf{QCMAH}$ algorithm distinguishing between the ``YES" distribution and the ``NO" distribution.

\begin{theorem}\label{thm:qcmah-close}
    The following holds for sufficiently large $\lambda$. Let $F:\{0,1\}^{[2^{\lambda^7}]}\to \{0,1,\perp\}$ be a partial function that can be computed by a circuit of size $s$ and depth $2d$, where each gate is either an OR gate (of unbounded fan-in), or a partial function that can be computed by a $T$-query quantum algorithm. Identify the set $\{0,1\}^{\lambda^7}$ with $[2^{\lambda^7}]$, and identify a function $f_\lambda:\{0,1\}^{\lambda^7}\to \{0,1\}$ with a binary string of length $2^{\lambda^7}$ (i.e.\ an element of $\{0,1\}^{[2^{\lambda^7}]}$). Then,
    $$\Pr_{f_\lambda\gets \mathscr D_{\text{YES}}}\left[F(f_\lambda)=1\right]+\Pr_{f_\lambda\gets \mathscr D_{\text{NO}}}\left[F(f_\lambda)=0\right]\leq 1+s(1000e^{-\lambda/100d}T^2)^{e^{\lambda/20d}}+e^{-\lambda/20d}\,.$$
\end{theorem}
\begin{proof}
    We construct the distribution of sets $D_{2d+1,0},\cdots,D_{2d+1,2d+1}$ as above. Let $t=e^{\lambda/20d}$. Then, by Lemma~\ref{lem:sampling_D_probability}, each $D_{2d+1,k}$ is $2t$-wise $e^{-\lambda/3(2d+1)}$-bounded. Thus, for a random $R:\{0,1\}^{\lambda^7}\to \{0,1\}$ and $D_H=D_{2d+1,2d+1}$, define the restriction $\rho$ as
    $$\rho(i)=\begin{cases}
        * & i\in D_H\\
        R(i) & i\notin D_H
    \end{cases}\,.$$
    We have, by Theorem~\ref{thm:qcmah-restriction},
    $$\Pr[\dt(F_\rho)\geq 1]\leq s(1000e^{-\lambda/100d}T^2)^{e^{\lambda/20d}}+e^{-\lambda/20d}\,.$$
    Now, for any fixed $R$ and $D_H$, if $\dt(F_\rho)=0$, it is impossible to have \emph{both} $F(R\lor \mathbf 1_{D_H})=1$ \emph{and} $F(R)=0$. This yields the desired theorem.
\end{proof}

\subsection{The separation}

Through Corollary~\ref{cor:qma-yes} and Lemma~\ref{lem:qma-no}, we have seen that a $\qma$ algorithm distinguishes the ``YES" distribution and the ``NO" distribution; and through Theorem~\ref{thm:qcmah-close}, we see that a $\qcmah$ algorithm cannot distinguish them. The next step will be to construct a separating language based on the distributions, which is standard in the literature, e.g.~\cite{RT19-raz-tal,AIK22,AB26}.

\begin{theorem}\label{thm:separation}
    There exist an oracle $\mathcal O$ and a decision problem $\mathcal L$, such that $\mathcal L\in \qma^{\mathcal O}$, and $\mathcal L\notin \qcmah^{\mathcal O}$.
\end{theorem}

\begin{proof}
    We construct the oracle as follows. For any $\lambda\in \mathbb N_+$,
    \begin{itemize}
        \item with probability $1/2$, we sample $f_\lambda:\{0,1\}^{\lambda^7}\to \{0,1\}$ from $\mathscr D_{\text{YES}}$, set $\mathcal O|_{\{0,1\}^{\lambda^7}}:=f_\lambda$, and put $1^\lambda\in \mathcal L$;
        \item with probability $1/2$, we sample $f_\lambda:\{0,1\}^{\lambda^7}\to \{0,1\}$ from $\mathscr D_{\text{NO}}$, set $\mathcal O|_{\{0,1\}^{\lambda^7}}:=f_\lambda$, and set $1^\lambda\notin \mathcal L$.
    \end{itemize}
    All other values of $\mathcal O$ not defined above are set to $0$, and all other strings not mentioned above are excluded from \(\mathcal L\).

    We first show that, with high probability, there exists a $\qma^{\mathcal O}$ algorithm that decides $\mathcal L$. In fact, we show that the algorithm $V$ defined in Algorithm~\ref{alg:QMA-ver} is a $\qma^{\mathcal O}$ verifier for $\mathcal L$ with high probability, after a suitable amplification. Indeed, by Corollary~\ref{cor:qma-yes} and Lemma~\ref{lem:qma-no} (where we choose $\eta=1/100$), for every sufficiently large $n$, with probability at least $1-2^{-n}$, the following holds.
    \begin{itemize}
        \item If $1^n\in \mathcal L$, there exists a quantum witness $\ket{\adv}$ such that $\Pr[V^{\mathcal O}(1^n,\ket{\adv})=1]\geq 99/100$.
        \item If $1^n\notin \mathcal L$, for all quantum witnesses $\ket{\adv}$, $\Pr[V^{\mathcal O}(1^n,\ket{\adv})=1]\leq 51/100$.
    \end{itemize}
    We choose a sufficiently large $n_0$. After a standard amplification, for all $n\geq n_0$, $V^{\mathcal O}$ is the desired $\qma^{\mathcal O}$ verifier, with probability at least
    $$\prod_{n\geq n_0}(1-2^{-n})\geq 0.9\,.$$
    And we may hardcode the language for all $n<n_0$, thus getting the $\qma^{\mathcal O}$ algorithm for $\mathcal L$.

    Next, we show that, with high probability, there does not exist a $\qcmah^{\mathcal O}_d$ algorithm that decides $\mathcal L$, for any $d\geq 1$.
    Let $A$ be any $\pqcmah_d^{\mathcal O}$ algorithm. By Proposition~\ref{prop:qcmah-circuits}, there exists a polynomial $q$, and for all sufficiently large $n$, there exists a circuit $C$ of size $2^{q(n)}$ and depth $2d$, where each gate is either an OR gate or a partial function that can be computed by a $q(n)$-query quantum algorithm, such that on input $\mathcal O_{\leq q(n)}$, the circuit $C$ outputs $A^{\mathcal O}(1^n)$. Note that the oracle value on each input length is chosen independently. Using Theorem~\ref{thm:qcmah-close}:
    \begin{align*}
        \Pr_{f_n\gets \mathscr D_{\text{YES}}}\left[A^{\mathcal O}(1^n)=1\right]+\Pr_{f_n\gets \mathscr D_{\text{NO}}}\left[A^{\mathcal O}(1^n)=0\right]\leq {}&1+2^{q(n)}(1000e^{-n/100d}q(n)^2)^{e^{n/20d}}+e^{-n/20d}\\
        \leq{}& 1+o(1)\,.
    \end{align*}
    In particular, by the construction of the oracle, $A^{\mathcal O}$ correctly decides $1^n$ with probability no more than $1/2+o(1)$, for sufficiently large $n$.

    By the independence of $\mathcal O$ on different input lengths, and the fact that $A$ can only ask queries of length at most $q(n)$ on input $1^n$, we get that there is an infinite sequence $n_1,n_2,\cdots$ such that
    $$\Pr[A^{\mathcal O}(1^{n_{i+1}})=\mathcal L(1^{n_{i+1}})\mid A^{\mathcal O}(1^{n_{1}})=\mathcal L(1^{n_{1}})\land \cdots \land A^{\mathcal O}(1^{n_{i}})=\mathcal L(1^{n_{i}})]\leq 0.51\,.$$
    Thus,
    $$\Pr[\forall n\in \mathbb N_+,\ A^{\mathcal O}(1^n)=\mathcal L(1^n)]\leq \prod_{n\in \{n_i\}_{i\geq 1}}0.51=0\,.$$
    And since there are countably many $\qcmah$ machines, we have that with probability $1$, no $\qcmah^{\mathcal O}$ algorithm decides $\mathcal L$.

    Overall, we get that with probability at least $0.9$ over the sampling of $\mathcal O$, $\mathcal L\in \qma^{\mathcal O}$ and $\mathcal L\notin \qcmah^{\mathcal O}$.
\end{proof}

%% file: Proof/collapse.tex
In this section, we leverage our definition of $\pqcmah$ to prove the following theorem.

\begin{theorem}\label{thm:collapse}
    There exists a classical oracle $\mathcal O$, such that
    $$\mathsf{PromiseP}^{\mathcal O}=\mathsf{PromiseQCMA}^{\mathcal O}=\pqcmah^{\mathcal O}\,,$$
    and
    $$\qcma^{\mathcal O}\subsetneq \qma^{\mathcal O}\,.$$
\end{theorem}

The same technique also solves the following open question in~\cite{AIK22}.

\begin{theorem}\label{thm:collapse-qma}
    There exists a classical oracle $\mathcal O$, such that
    $$\mathsf{PromiseP}^{\mathcal O}=\mathsf{PromiseQMA}^{\mathcal O}=\mathsf{PromiseQMAH}^{\mathcal O}\,,$$
    and
    $$\qma^{\mathcal O}\subsetneq \mathsf{PP}^{\mathcal O}\,.$$
\end{theorem}

As discussed in~\cite{AIK22}, the usual technique does not directly apply, because one must assign a value to every out-of-promise input of a promise language defined by a $\qcma$ or $\qma$ machine. We address this issue by choosing these values during diagonalization. Using the convention in Definition~\ref{def:partial-func-extension} for $\pqcmah$ (resp.\ $\pqmah$), our separation result indicates that there exists a consistent completion that precludes the current machine. We choose such a completion and use it to assign the corresponding out-of-promise oracle values.

Due to the similarity between Theorem~\ref{thm:collapse} and Theorem~\ref{thm:collapse-qma}, we give only the proof of the former in detail.

The oracle in Theorem~\ref{thm:collapse} will consist of two parts. One part $A=\{A_n\}$ will store instances drawn from the ``YES'' and ``NO'' distributions in Definition~\ref{def:separating-problem}, and another part $B=\{B_n\}$ will store the answers of $\pqcma$ verifiers relative to the oracle.

The proof of Theorem~\ref{thm:collapse} proceeds as follows. First, we introduce a definition of \emph{acceptable} oracles. An acceptable oracle construction collapses $\pqcma$ to $\mathsf{PromiseP}$. We then analyze $\mathsf{PromiseP}^{\mathcal O}$ machines, which semantically correspond to $\pqcmah^A$ machines under acceptable assignments of $B$. Finally, we apply a diagonalization argument similar to that in Theorem~\ref{thm:separation} to separate $\mathsf{PromiseP}^{\mathcal O}$ from $\pqma^{\mathcal O}$.

We now start to construct the oracles. Similarly to~\cite{AIK22}, we first present the following definitions.

\begin{definition}
    For $t\in \mathbb N_+$, let $S_t$ be the set of all strings $\langle V\rangle$ satisfying the following conditions.
    \begin{enumerate}
        \item $\langle V\rangle$ takes exactly $t$ bits to specify.
        \item $\langle V\rangle $ specifies a quantum oracle circuit $V$ with the following syntax: the circuit has at most $t$ gates, accepts a string $w$ of at most $t$ bits, and queries the oracle only on length $\leq \sqrt{t}$.
    \end{enumerate}
\end{definition}

It is clear that membership in $S_t$ can be verified in polynomial time.

We start by describing part $B$ of our oracle. However, we do not determine the complete truth table of $B$ right now; instead, we specify what condition makes $B$ acceptable in our following analysis.

\begin{definition}\label{def:acceptable}
    Let $\mathcal O=(A,B)$ be an oracle. We call $\mathcal O_{\leq n}$ \emph{acceptable}, if for all $t\leq n$, the following hold.
    \begin{itemize}
        \item For every $v\in \{0,1\}^t\backslash S_t$, $B(v)=0$.
        \item For every $\langle V\rangle \in S_t$, we have
        $$\max_{w}\Pr[V^{\mathcal O}(w)=1]\geq \frac23\implies B(\langle V\rangle)=1\,,$$
        $$\max_{w}\Pr[V^{\mathcal O}(w)=1]\leq \frac13\implies B(\langle V\rangle)=0\,.$$
        Notice that $\max_{w}\Pr[V^{\mathcal O}(w)=1]$ depends only on $\mathcal O_{\leq \sqrt{t}}$.
    \end{itemize}

    Call $\mathcal O$ acceptable, if for all $n$, $\mathcal O_{\leq n}$ is acceptable.
\end{definition}

Note that an acceptable oracle can be arbitrarily set on $\langle V\rangle$ if $\max_{w}\Pr[V^{\mathcal O}(w)=1]$ is between $1/3$ and $2/3$. Moreover, this arbitrariness can influence the values of the oracle, since machines sitting in $S_{t'}$ with larger $t'$ may query $B(\langle V\rangle)$, and their behavior may depend on that oracle value. Nevertheless, given an acceptable $\mathcal O_{\leq n}$, for all $n'>n$, there always exists an acceptable extension $\mathcal O_{\leq n'}$: simply decide the values of $\mathcal O$ from shorter inputs to longer inputs, and choose arbitrary values wherever they are not determined by Definition~\ref{def:acceptable}.

We prove that this notion of acceptability suffices to collapse $\pqcma$ to $\mathsf{PromiseP}$, while giving the freedom of choosing the out-of-promise values in an adversarial manner.

\begin{proposition}\label{prop:acceptable-to-collapse}
    For all acceptable oracles $\mathcal O$,
    $$\mathsf{PromiseP}^{\mathcal O}=\mathsf{PromiseQCMA}^{\mathcal O}=\pqcmah^{\mathcal O}\,.$$
\end{proposition}

\begin{proof}
    Let $\mathcal L$ be a $\pqcma^{\mathcal O}$ language. By definition, given $x\in \{0,1\}^\lambda$, there exists a polynomial time Turing machine $M$ that on input $1^\lambda$, outputs a description of some circuit $\mathcal A$, such that
    $$x\in \mathcal L_{\text{YES}}\implies \max_w \Pr[\mathcal A^{\mathcal O}(x,w)=1]\geq \frac23\,,$$
    $$x\in \mathcal L_{\text{NO}}\implies \max_w \Pr[\mathcal A^{\mathcal O}(x,w)=1]\leq \frac13\,.$$
    We hardcode the string $x$ into the circuit, getting a circuit $V$ that queries $\mathcal O$, and behaves exactly as $\mathcal A^{\mathcal O}(x,\cdot)$. Moreover, we may pad the description of $V$ such that it lies in $S_t$ for some $t$. And thus, we get an $\mathcal O$-oracle polynomial time Turing machine that acts as follows: it receives a string $x$, constructs the description $\langle V\rangle$ of the circuit $V$, and queries $B(\langle V\rangle)$. This Turing machine clearly decides the promise problem $\mathcal L$, and thus $\pqcma^{\mathcal O}\subseteq \mathsf{PromiseP}^{\mathcal O}$.

    Since $ \mathsf{PromiseP}^{\mathcal O}\subseteq\pqcma^{\mathcal O}$, the equality holds. By the definition of $\pqcmah$, it also collapses to $\mathsf{PromiseP}$ under the oracle $\mathcal O$.
\end{proof}

Next, we prove the following, which essentially states that, given $t'\leq \poly(t)$,
\begin{enumerate}
    \item each bit in $B_{t'}$ corresponds to a constant-depth circuit of OR gates and quantum query gates whose input does not depend on $B_i$ for $i> t$, and
    \item the consistent evaluation (Definition~\ref{def:partial-func-extension}) of this circuit corresponds to an \emph{acceptable} choice of $\mathcal O_{\leq t'}$ (Definition~\ref{def:acceptable}).
\end{enumerate}

\begin{proposition}\label{prop:acceptable-to-consistent}
    Let $t\in \mathbb N_+$ be sufficiently large. Fix $d\in \mathbb N_+$, and let $t'\leq t^{2^d}$. For each $\langle V\rangle \in S_{t'}$, there exists a circuit $C$ of size $2^{t'+1}$ and depth $2d$, such that
    \begin{itemize}
        \item Each gate in $C$ is either an OR gate, or a partial function that can be computed by a $t'$-query quantum algorithm;
        \item $C$ takes $A_{\leq t'}$ and $B_{\leq t}$ as input; and
        \item Fix $A_{\leq t'}$ and $B_{\leq t}$ such that $\mathcal O_{\leq t}$ is acceptable. For each consistent completion $\widehat C$ of $C$, we can choose $\mathcal O_{\leq t'}$ (that is, choose $B_{t+1},\cdots,B_{t'}$), such that
        \begin{itemize}
            \item $\mathcal O_{\leq t'}$ is acceptable, and
            \item $B(\langle V\rangle)$ equals the output of $\widehat C$.
        \end{itemize}
    \end{itemize}
\end{proposition}
\begin{proof}
    This proof is similar to~\cite[Lemma~35]{AIK22}.

    More formally, we prove this proposition by induction on $d$.
    For the base case $d=1$, we have $t'\leq t^2$. By the definition of $S_{t'}$, $V$ is a quantum oracle circuit with at most $t'$ gates, accepts a string $w$ of at most $t'$ bits, and queries the oracle only on length $\leq t$. This means that, for each witness $w$, $V^{(\cdot)}(w)$ corresponds to a function of $A_{\leq t'}$ and $B_{\leq t}$ that can be computed by a $t'$-query quantum algorithm. Maximizing the probability over $w$ can be replaced by an OR gate over all witnesses, and thus we convert $V$ into a circuit of size $2^{t'}+1\leq 2^{1+t'}$.

    For the inductive step, let $d\geq 2$. As in the base case above, we use the definition of $S_{t'}$ to obtain a circuit of depth $2$ and size $2^{t'}+1$ that takes as input $\mathcal O_{\leq \sqrt{t'}}$. We then use the inductive hypothesis to expand the bits in $B_{t+1},\cdots,B_{\sqrt{t'}}$ into a circuit of depth at most $2d$ and size at most
    $$2^{t'}+1+\sum_{i=t+1}^{\sqrt{t'}}2^i\cdot 2^{1+i}\leq 2^{1+t'}\,.$$
    Crucially, this circuit should use a single gate to compute each intermediate value of the $B$ oracle, and all subsequent evaluations should use the same result from this gate. Thus, each consistent completion of this circuit corresponds exactly to the selection of the outputs of these intermediate gates. Hence, each consistent completion yields an acceptable choice of the oracle values of $B$.
\end{proof}

We have now built all ingredients for Theorem~\ref{thm:collapse}, and the rest is similar to the proof of Theorem~\ref{thm:separation}.

\begin{proof}[Proof of Theorem~\ref{thm:collapse}]
    We will construct an acceptable $\mathcal O$ and a language $\mathcal L\in \qma^{\mathcal O}\backslash \mathsf{P}^{\mathcal O}$. Together with Proposition~\ref{prop:acceptable-to-collapse}, this construction proves the desired theorem.
    
    Let $M_1,M_2,\cdots$ be an enumeration of polynomial time oracle Turing machines, such that each $M_\tau$ has a description of length at most $\tau$, and runs in at most $n^\tau$ steps when the input length is $n$. We construct an efficiently computable sequence $n_0,n_1,n_2,\cdots$ satisfying the following.
    \begin{itemize}
        \item $n_0$ is a sufficiently large fixed integer.
        \item For each $i$, $n_i\geq (n_{i-1})^{4i+100}$.
        \item $n_i\geq 2^i$.
    \end{itemize}
    We choose an arbitrary acceptable $\mathcal O_{\leq \sqrt{n_1}}$, and set $\mathcal L_{\leq n_0}=0$. Now, for each $i\in \mathbb N_+$, we construct $\mathcal O_{\leq n_i}$ as follows.

    Assume $\mathcal O_{\leq \sqrt{n_i}}$ is already fixed and acceptable, and $\mathcal L_{\leq n_{i-1}}$ is already fixed. $M_i$ has a description of length at most $i$, and $M_i^{(\cdot)}(1^{n_i})$ runs in $(n_i)^{i}$ steps, and queries the oracle of length at most $(n_i)^i$. We pad the description of $M_i^{(\cdot)}(1^{n_i})$ to get $v_i\in S_{(n_i)^{2i}}$, where $v_i$ describes the oracle circuit $M_i^{(\cdot)}(1^{n_i})$ that takes an empty string as input, and queries the oracle only on length at most $(n_i)^i$.

    Let $t=\sqrt{n_i}$ and $t'=(n_i)^{2i}$.
    We construct the circuit $C$ corresponding to $v_i$ in Proposition~\ref{prop:acceptable-to-consistent}, which is of size $2^{t'+1}$ and depth at most $2\log (4i)\leq 4i$, taking $A_{\leq t'}$ and $B_{\leq t}$ as input. We then hardcode ${\mathcal O}_{\leq t}$ into the circuit (since it is already determined). And moreover, we set $A_{t+1},\cdots,A_{t'}$ to $0$, except for $A_{n_i^7}$, which yields a circuit of size at most $2^{t'+2}$, which we still call $C$ by an abuse of notation.
    
    Let $f_{n_i}:\{0,1\}^{n_i^7}\to \{0,1\}$ be a boolean function. Theorem~\ref{thm:qcmah-close} yields the following:
    \begin{align*}
        &\Pr_{f_{n_i}\gets \mathscr D_{\text{YES}}}[C(f_{n_i})=1]+\Pr_{f_{n_i}\gets \mathscr D_{\text{NO}}}[C(f_{n_i})=0]\\
        \leq{}& 1+2^{(n_i)^{2i}+2}(1000e^{-n_i/400i}(n_i)^{4i})^{e^{n_i/80i}}+e^{-n_i/80i}\\
        \leq{} & 1+o(1)\,.
    \end{align*}
    In particular, one of the following events has probability at least $0.49$.
    \begin{enumerate}
        \item Sample $f_{n_i}$ from $\mathscr D_{\text{YES}}$. There exists a consistent completion $\widehat C$ of $C$, such that $\widehat C(f_{n_i})=0$.
        \item Sample $f_{n_i}$ from $\mathscr D_{\text{NO}}$. There exists a consistent completion $\widehat C$ of $C$, such that $\widehat C(f_{n_i})=1$.
    \end{enumerate}
    By Proposition~\ref{prop:acceptable-to-consistent}, one of the following holds with probability at least $0.49$.
    \begin{enumerate}
        \item Sample $A_{n_i^7}$ from $\mathscr D_{\text{YES}}$. We can choose an acceptable $\mathcal O_{\leq (n_i)^{2i}}$ such that $B(v_i)=0$. In particular, the acceptability of $\mathcal O_{\leq (n_i)^{2i}}$ requires that $M_i^{\mathcal O}(1^{n_i})=0$.
        \item Sample $A_{n_i^7}$ from $\mathscr D_{\text{NO}}$. We can choose an acceptable $\mathcal O_{\leq (n_i)^{2i}}$ such that $B(v_i)=1$. In particular, the acceptability of $\mathcal O_{\leq (n_i)^{2i}}$ requires that $M_i^{\mathcal O}(1^{n_i})=1$.
    \end{enumerate}
    However, by Corollary~\ref{cor:qma-yes} and Lemma~\ref{lem:qma-no}, there is a $\qma^{\mathcal O}$ algorithm $V^{\mathcal O}$ (which is independent of $i$) such that \emph{both} of the following hold with probability at least $1-o(1)$.
    \begin{enumerate}
        \item If $A_{n_i^7}$ is sampled from $\mathscr D_{\text{YES}}$, then $V^{\mathcal O}(1^{n_i})=1$.
        \item If $A_{n_i^7}$ is sampled from $\mathscr D_{\text{NO}}$, then $V^{\mathcal O}(1^{n_i})=0$.
    \end{enumerate}

    Thus, with probability at least $0.48$, one of the following holds.
    \begin{enumerate}
        \item Sample $A_{n_i^7}$ from $\mathscr D_{\text{YES}}$, and put $1^{n_i}\in \mathcal L$. There exists an acceptable $\mathcal O_{\leq (n_i)^{2i}}$ such that $M_i^{\mathcal O}(1^{n_i})=0$, and $V^{\mathcal O}(1^{n_i})=1$.
        \item Sample $A_{n_i^7}$ from $\mathscr D_{\text{NO}}$, and set $1^{n_i}\notin \mathcal L$. There exists an acceptable $\mathcal O_{\leq (n_i)^{2i}}$ such that $M_i^{\mathcal O}(1^{n_i})=1$, and $V^{\mathcal O}(1^{n_i})=0$.
    \end{enumerate}

    In particular, given an acceptable $\mathcal O_{\leq \sqrt{n_i}}$, there exists an acceptable extension $\mathcal O_{\leq (n_i)^{2i}}$ and the corresponding choice of $\mathcal L(1^{n_i})$, such that $M_i^{\mathcal O}$ cannot decide $\mathcal L$ on input $1^{n_i}$, but $V^{\mathcal O}$ can.

    Since $(n_i)^{2i}\leq \sqrt{n_{i+1}}$, we may carry out the induction on $i$ and thus get a construction of an acceptable oracle $\mathcal O$ and the language $\mathcal L$, such that $\mathcal L\notin \mathsf{P}^{\mathcal O}$, but the $\qma^{\mathcal O}$ algorithm $V^{\mathcal O}$ decides $\mathcal L$ on every input $1^{n_i}$ with $i\geq 1$. Since the sequence $n_i$ is efficiently computable, $\mathcal L\in \qma^{\mathcal O}$. Proposition~\ref{prop:acceptable-to-collapse} implies that $\mathsf{PromiseP}^{\mathcal O}=\mathsf{PromiseQCMA}^{\mathcal O}$, and in particular, $\mathsf{P}^{\mathcal O}=\qcma^{\mathcal O}$, which finishes the proof.
\end{proof}

The proof of Theorem~\ref{thm:collapse-qma} is similar to that of Theorem~\ref{thm:collapse}, so we only present a sketch of the proof.

\begin{proof}[Proof of Theorem~\ref{thm:collapse-qma}]
    The theorem follows from the same argument as in the proof of Theorem~\ref{thm:collapse}, except that Theorem~\ref{thm:qcmah-close} is replaced with the following lemma.
    \begin{lemma}[Implicit in~\cite{AIK22}, Corollary 70]\label{lem:collapse-qma-close}
    Let $F:\{0,1\}^N\to \{0,1,\perp\}$ be computed by a circuit $C$ of size $s=\mathrm{quasipoly}(N)$ and depth $d=O(1)$, where each gate has $\qma$ query complexity at most $R\leq \poly\log(N)$, and \emph{the evaluation of $C$ follows Definition~\ref{def:partial-func-extension}}. Then for any $\epsilon\geq 1/\mathrm{quasipoly}(N)$ and sufficiently large $N$,
    $$\Pr_{x:\operatorname{parity}(x)=1}[F(x)=1]+\Pr_{x:\operatorname{parity}(x)=0}[F(x)=0]\leq 1+\epsilon\,.$$
\end{lemma}
Using the same diagonalization technique, we get a classical oracle relative to which $\mathsf{PromiseP}=\mathsf{PromiseQMA}$ while $\mathsf{\oplus P}$ remains strictly larger than $\mathsf{P}$, and thus $\mathsf{PP}$ is strictly larger than $\mathsf{P}$.
\end{proof}

%% file: Appendix/blackbox.tex
This appendix lists definitions that we use as black boxes in our proof, including the definition of the multiplicity code and the biased Yamakawa--Zhandry algorithm. We include them here for completeness.

\subsection{The multiplicity code}

\begin{definition}[Formal and Hasse derivatives]
Let
\(
    f(X)=\sum_{j=0}^{n} a_jX^j \in \mathbb{F}_q[X]
\)
be a univariate polynomial over $\mathbb{F}_q$. We define the $i$-th
formal and Hasse derivatives of $f(X)$ as the linear operators which
take $f(X)$ to the polynomials
\[
    f^{[i]}(X)
    = \sum_{j=i}^{n} \frac{j!}{(j-i)!}\,a_jX^{j-i}
    \qquad\text{and}\qquad
    f^{(i)}(X)
    = \sum_{j=i}^{n} \binom{j}{i}a_jX^{j-i},
\]
respectively. Note that $f^{[i]}(X)=i!\,f^{(i)}(X)$ for all $i$ and $f$.
\end{definition}

\begin{definition}[Univariate multiplicity codes]\label{def:mult-code}
Let $\mathbb{F}_q$ be a finite field and let $s$ be a positive integer.
Let $\alpha_1,\ldots,\alpha_n$ be distinct points in $\mathbb{F}_q$,
and let $k<sn$ be a positive integer. The univariate multiplicity code
\(
    \mult_{s,\mathbb{F}_q}
    (\alpha_1,\ldots,\alpha_n;k)
\)
is the code over the alphabet $\Sigma=\mathbb{F}_q^s$ of length $n$
which associates each polynomial $f(X)\in\mathbb{F}_q[X]_{<k}$ to
the codeword $c\in\Sigma^n$ such that, for $i\in[n]$,
\[
    c_i
    = \bigl(
        f^{(0)}(\alpha_i),
        f^{(1)}(\alpha_i),
        \ldots,
        f^{(s-1)}(\alpha_i)
      \bigr),
\]
where $f^{(j)}$ is the $j$th Hasse derivative of $f$. Define
\[
    \mult_{s,\mathbb{F}_q,k}
    := \mult_{s,\mathbb{F}_q}(1,\ldots,q;k)\,.
\]
\end{definition}

\subsection{The biased Yamakawa--Zhandry algorithm}

For each parameter $\lambda\in \mathbb N_+$, let $C_\lambda=\mult_{s,\mathbb F_q,k}$ be the multiplicity code defined as in Definition~\ref{def:mult-code}, where $k=\lambda^3$, $q$ is the smallest prime greater than $\lambda^5$, and $s=\lambda$. In particular, $\lambda^5<q\leq 2\lambda^5$. Define $\Sigma=\mathbb F_q^s$.

Define $\mathsf{Dec}_{C_\lambda^\perp}$ as the decoder for the dual code $C_\lambda^\perp$.

Define the Hilbert space $\mathcal H=\mathbb C^{q^s}$, where the basis is labeled by elements in $\Sigma$. Let $\{\mathsf{A}_{i,c}\}_{i\in [q],c\in \{0,1\}}$ be registers representing $\mathcal H$. Let $\mathsf B$ be a register representing $\mathcal H^{\otimes q}$.

\begin{algorithm}[htbp]
    \caption{The biased Yamakawa--Zhandry algorithm $\mathsf{BiasedYZ}$}
    \label{alg:biasedYZ}
    \begin{algorithmic}
        \State \textbf{Input:} A state $\ket{\adv}$ on registers
        $\{\mathsf{A}_{i,c}\}_{i\in[q],\,c\in\{0,1\}}$ and $\mathsf{B}$,
        and a string $(b_1,\ldots,b_\lambda)\in\{0,1\}^{\lambda}$.
        \State Set $b_i:=0$ for $\lambda<i\leq q$
        \State Define the register $\mathsf{A}=(\mathsf{A}_{1,b_1},\ldots,\mathsf{A}_{q,b_q})$
        \State Apply $\mathsf{QFT}_{\Sigma}^{\otimes q}$ to $\mathsf{A}$
        \State Apply $\mathsf{QFT}_{\Sigma}^{\otimes q}$ to $\mathsf{B}$
        \State Apply the map $\ket{e}\ket{u} \mapsto \ket{e+u}\ket{u}$ to $(\mathsf{A},\mathsf{B})$
        \State Apply the map $\ket{z}\ket{u} \mapsto
            \ket{z}\ket{u-\mathsf{Dec}_{C_\lambda^\perp}(z)}$ to $(\mathsf A,\mathsf B)$
        \State Apply $\left(\mathsf{QFT}_{\Sigma}^{-1}\right)^{\otimes q}$ to $\mathsf{A}$
        \State Measure $\mathsf{A}$
        in the computational basis to get the outcome $v\in\Sigma^q$
        \State \textbf{Output:} $v$
    \end{algorithmic}
\end{algorithm}

The honest witness $\ket{\adv_H}$ in Lemma~\ref{lem:Biased-YZ-succeeding} is as follows.
Let
$S_{i,c}:=\{a\in\Sigma:H(i,a)=c\}$.
When all these sets are nonempty, $\ket{\adv_H}$ is a product state in which
each $\mathsf{A}_{i,c}$ contains the uniform superposition over $S_{i,c}$
and $\mathsf{B}$ contains the uniform superposition over $C_\lambda$.
If any $S_{i,c}$ is empty, we may take $\ket{\adv_H}$ to be arbitrary.

%% file: Appendix/AI-disclaimer.tex
The author believes that both the human author and the AI played indispensable roles in this work. This section gives a more detailed account of their respective contributions. Unless otherwise noted, ``the AI'' refers to GPT-6 Astra.

\paragraph{Initial proof of the separation.}
The author originated the idea of separating $\qma$ from the $\qcma$ hierarchy while attempting to construct a classical oracle separation between Microcrypt primitives (see Section~\ref{sec:intro}). When given this task, GPT-6 Astra constructed an oracle separating $\qma$ from an alternative complexity class, denoted by $\qcmah_{\text{robust}}$ in Remark~\ref{rmk:qcmah_alt}. The AI produced nearly the entire proof in a single attempt, which took approximately 10 hours using GPT-6 Astra Ultra in fast mode.

The initial proof was complicated and used an oracle substantially different from the one presented in this paper. After reading it, the author suggested several simplifications to the AI, including the oracle construction in Section~\ref{sec:proof-main}. Using this oracle, the AI then produced a proof separating $\qma$ from $\qcmah_{\text{robust}}$. The proof used the same technique as the first attempt. It used Fourier analysis to argue that changing the oracle from the ``NO'' distribution to the ``YES'' distribution had little effect on the behavior of a sampler similar to the algorithm $\mathsf{Guesser}$ in~\cite[Section~6.2]{BHV26}.

\paragraph{The proof via the switching lemma.}
While trying to simplify the proof further and generalize it to $\qcmah$, the author identified a form of switching lemma as the crucial idea in an AI-generated proof. In studying the argument, the author isolated a switching lemma sharper than the one in~\cite{AB26} and recognized that it deserves a separate treatment. The parameters in the AI's version still depended on the number of input variables, and its proof differed substantially from the one presented in Section~\ref{sec:switching}.

Independently of the AI, the author first proved the switching lemma for quantum algorithms (Corollary~\ref{cor:quantum_switching}) using query weights, inspired by earlier work~\cite{FK18,BHNZ26,BHV26}. After comparing this argument with~\cite{AB26}, the author realized that it also applied to polynomials, yielding Theorem~\ref{thm:quantum_switching}. The author's proof is arguably simpler than the AI's version and much easier for a human reader to follow. The author also derived a version for DNFs and later learned that it was already known (Theorem~\ref{thm:kelley-switching}).

Once the author had proved Corollary~\ref{cor:quantum_switching}, the argument in Section~\ref{sec:proof-main} was straightforward. The author worked it out independently, except for Section~\ref{sec:proof-main-qma}, whose argument was already known to the author through earlier interactions with the AI.

\paragraph{The collapse results.}
The human author developed the entire proof sketch in Section~\ref{sec:collapse}, with minimal help from the AI. However, the author also found that, with suitable guidance in the prompt, GPT-6 Astra Ultra in fast mode could prove the theorems in a single attempt taking approximately 15 minutes.

\paragraph{Proof audit.}
After completing the proof, the author asked the AI to audit it. The AI identified a subtlety in Proposition~\ref{prop:qcmah-circuits}, which originates from the circuit formulation of~\cite{AB26}. Without a correction, the argument would establish the separation only for the class $\qcmah_{\text{robust}}$ discussed in Remark~\ref{rmk:qcmah_alt}.\footnote{As discussed in Remark~\ref{rmk:conventions}, there are two conventions for evaluating a circuit whose gates compute partial functions. One is used in~\cite{AIK22}, and the other is the consistent completion convention (Definition~\ref{def:partial-func-extension}) used in this paper. The latter corresponds to our definition of $\pqcmah$ (Definition~\ref{def:qcmah}). The proof in~\cite{AB26} does not explicitly specify which convention it uses but remains correct once this distinction is taken into account.} The AI suggested the correction presented in this paper.

\paragraph{Summary.}
\emph{The human author developed the proof presented in this paper independently, apart from the AI contributions described above}. In particular, the human author independently derived both the polynomial switching lemma (Theorem~\ref{thm:quantum_switching}) and the $\qcmah$ lower bound obtained using that lemma.

The human author wrote the entire manuscript, with the AI's help in polishing. The AI also assisted with parts of the literature review, especially the earlier work on bounded polynomials, with which the human author was less familiar.